\documentclass[floats,floatfix,showpacs,amssymb,prd,twocolumn,superscriptaddress,nofootinbib]{revtex4-1}

\usepackage{amssymb,amsmath,verbatim,mathtools,needspace,enumitem,etoolbox,graphicx,physics,microtype,afterpage}

\usepackage[dvipsnames, usenames]{xcolor}

\definecolor{linkcolor}{rgb}{0.0,0.3,0.5}
\usepackage[unicode, colorlinks=true, linkcolor=linkcolor, citecolor=linkcolor, filecolor=linkcolor,urlcolor=linkcolor, pdfusetitle]{hyperref}
\usepackage[all]{hypcap}
\usepackage[T1]{fontenc}
\usepackage[utf8]{inputenc}
\usepackage{tabularx}
\allowdisplaybreaks
\renewcommand{\arraystretch}{1.4}
\newcommand{\ssim}{\mathchar"5218\relax\,}

\def\apj{\rm{ApJ}}
\def\apjl{\rm{ApJ}}
\def\apjs{\rm{ApJS}}
\def\aap{\rm{A\&A}}
\def\aapr{\rm{A\&A~Rev.}}

\def\mnras{\rm{MNRAS}}
\def\prd{\rm{PRD}}
\def\prl{\rm{PRL}}

\def\nat{\rm{Nature}}

\def\cqg{\rm{CQG}}
\def\lrr{\rm{LRR}}

\newcommand{\caltech}{\affiliation{TAPIR 350-17, California Institute of Technology, 1200 E California Boulevard, Pasadena, CA 91125, USA}}
\newcommand{\rochester}{\affiliation{Center for Computational Relativity and Gravitation, Rochester Institute of Technology,
85 Lomb Memorial Drive, Rochester, NY 14623, USA}}

\newcommand{\jhu}{\affiliation{Department of Physics and Astronomy, Johns Hopkins University, 3400 N. Charles
Street, Baltimore, MD 21218, USA}}

\newcommand{\warsaw}{\affiliation{Nicolaus Copernicus Astronomical Center, Polish Academy of Sciences, ul. Bartycka 18, 00-716 Warsaw, Poland}}
\newcommand{\bham}{\affiliation{School of Physics and Astronomy and Institute for Gravitational Wave Astronomy, University of Birmingham, Birmingham, B15 2TT, UK}}

\begin{document}

\title{Multiband gravitational-wave event rates and stellar physics}

\author{Davide Gerosa}
\thanks{Einstein Fellow}
\email{dgerosa@caltech.edu}
\caltech \bham
\author{Sizheng Ma} \caltech
\author{Kaze W.K. Wong} \jhu
\author{Emanuele Berti} \jhu
\author{Richard O'Shaughnessy} \rochester
\author{Yanbei Chen} \caltech
\author{Krzysztof Belczynski} \warsaw

\date{\today}

\begin{abstract}

Joint gravitational-wave detections of stellar-mass black-hole binaries by ground- and space-based observatories will provide unprecedented opportunities for fundamental physics and astronomy. We present a semianalytic method to estimate multiband event rates by combining selection effects of ground-based interferometers (like LIGO/Virgo) and space missions (like LISA). %
We forecast the expected number of multiband detections first by using information from current LIGO/Virgo data, and then through population synthesis simulations of binary stars. We estimate that few to tens of LISA detections can be used to predict mergers detectable on the ground. Conversely, hundreds of events could potentially be extracted from the LISA data stream using prior information from ground detections. In general, the merger signal of binaries observable by LISA is strong enough to be unambiguously identified by both current and future ground-based detectors. Therefore third-generation detectors will not increase the number of multiband detections compared to LIGO/Virgo.  We use population synthesis simulations of isolated binary stars to explore some of the stellar physics that could be constrained with multiband events, and we show that specific formation pathways 
might
 be overrepresented in multiband events compared to ground-only detections.

\end{abstract}

\maketitle

\section{Science with multiband events}

The profound implications of LIGO's revolutionary discoveries for the LISA mission became clear soon after the first gravitational wave (GW) detection~\cite{2016PhRvL.116w1102S}. Black hole (BH) binaries of masses comparable to GW150914 merge at frequencies of $\ssim100\,{\rm Hz}$, where ground-based interferometers are most sensitive, but their mHz emission from the early inspiral is strong enough to be observed by LISA. LISA observations of the inspiral could provide good estimates of binary parameters such as the merger time and sky location, thus serving as forewarnings for GW merger observations from the ground and (possibly) electromagnetic counterparts.

\emph{Multiband} GW astronomy is now an important part of the LISA science case~\cite{2017arXiv170200786A}. %
Many authors have argued that LISA could make significant contributions to our understanding of stellar-mass BH astrophysics~\cite{2010ApJ...725..816B,2016ApJ...830L..18B,2016PhRvD..94f4020N,2017MNRAS.465.4375N,2018arXiv180208654S, 2019PhRvD..99f3003K,2019PhRvD..99f3006S,2018MNRAS.481.4775D,2018MNRAS.481.5445S,2018arXiv180208654S,2018PhRvL.120s1103K,2018arXiv180505335R}. One reason is that LISA could potentially measure binary properties that may not be accessible from the ground.
For instance, eccentricity is a common signature of dynamical formation channels. Eccentricity in the LIGO band is typically expected to be too low to be detectable (see e.g.~\cite{2018PhRvD..98h3028L}), because eccentric binaries quickly circularize under gravitational radiation reaction. However the eccentricity of dynamically formed binaries may be large enough to be measurable by LISA, giving us important clues about their formation history.

Joint LIGO-LISA detections also open up the possibility to perform new and more powerful tests of general relativity. For instance, some theories of gravity predict additional GW emission channels
~\cite{2015CQGra..32x3001B}. The combined analysis of LISA inspirals and ground-based mergers will put extremely stringent constraints on, e.g., dipolar radiation~\cite{2016PhRvL.116x1104B}. GW cosmology will also improve with mHz detections of stellar-mass BHs, because LISA's sky-localization properties  make them unique standard sirens in the local Universe~\cite{2017PhRvD..95h3525K,2018MNRAS.475.3485D}.

The prospect of multiband GW astronomy led to a flourishing of new data analysis and experimental ideas.
Advanced warning information on the merger time might allow ground-based operations to be specifically adjusted. For example, by ensuring all the detectors on the ground are taking data at the right time -- or perhaps even by tuning their optical properties~\cite{2018arXiv180700075T} -- we could achieve a qualitatively better characterization of specific ``golden'' sources. Data streams from ground- and space-based detector networks can also be combined to enrich the scientific payoff of {\em both} experiments. Binary properties measured with LISA could be used as a prior to improve ground-based parameter estimation pipelines~\cite{2016PhRvL.117e1102V}. Vice versa, ground-based detections can be exploited to revisit past LISA data looking for coincident triggers~\cite{2018PhRvL.121y1102W}. The effectiveness of waveform templates to characterize stellar-mass BHs with LISA is also being actively investigated~\cite{2018arXiv181101805M}.

Whether or not LISA will be able to deliver such revolutionary science crucially depends on the expected rates of multiband BH events~\cite{2016PhRvL.116w1102S,2017JPhCS.840a2018S,2016MNRAS.462.2177K}. The scope of this paper is twofold: 
\begin{enumerate}
\item We present a new procedure to convert merger rates measured by (or predicted for) ground-based detectors into multiband event rates. Our method relies on estimating the ``effective time window'' in which sources remain %
visible by LISA.
\item We make use of this method to explore the physics of massive binary stars that can potentially be uncovered with multiband GW detections. We first present ``model-agnostic'' estimates based only on current observational bounds from LIGO/Virgo. We then compute rates using population synthesis models of isolated BH binaries.
\end{enumerate}

This paper is organized as follows. In Sec.~\ref{methodrates} we describe our method to compute event rates. Section~\ref{datapredictions} translates the current event rate measured by LIGO/Virgo into predictions for LISA. In Sec.~\ref{spopssection} we apply our findings to state-of-the-art population synthesis simulations of merging stellar-mass BHs. In Sec.~\ref{conclusions} we compare against previous estimates and discuss topics that should be addressed in future research. To improve readability, we present some of our results in Appendix~\ref{suppmat} (these include, in particular, long tables listing multiband rates for different assumptions and population synthesis models). In Appendix~\ref{zhorsec} we discuss the horizon redshift of ground- and space-based detectors.  Throughout the paper we use geometrical units ($G=c=1$) and we use values of the cosmological parameters drawn from Ref.~\cite{2016A&A...594A..13P}.

\section{Combining event rates}
\label{methodrates}

An astrophysical BH binary is described by a set of  intrinsic parameters $\lambda$.
Depending on the model/measurements available, these might include quantities like source-frame masses $m_1>m_2$, spin vectors $\boldsymbol{\chi}_1$ and $\boldsymbol{\chi}_2$ and binary eccentricity $e$. 
Sources are located at a cosmological redshift $z$, where the intrinsic merger rate is $\mathcal R(z)$ (this typically measured in units of Gpc$^{-3}$ yr$^{-1}$).

Let us now examine selection effects for ground- and space-based detectors separately.

\subsection{Detection rate from the ground}

Ground-based detection rates $r$ (in units of yr$^{-1}$) are related to the intrinsic merger rate via
\begin{align}
r_{\rm ground}= \iint dz d\lambda\;  \mathcal{R}(z)\;p(\lambda) \frac{d V_c(z)}{dz} \frac{1}{1+z}\; p_{\rm det}(\lambda, z)\,,
\label{rground}
\end{align}
where ${d V_c(z)}/{dz}$ is the shell of comoving volume at redshift $z$, the factor $1/(1+z)$ accounts for the Universe's expansion between emission and detection, $p(\lambda)$ is the probability density function of the intrinsic parameters, and $0\leq p_{\rm det}(\lambda,z)\leq 1$ is a detection probability.

The accurate estimate of the detector's sensitivity volume is a crucial element in current LIGO/Virgo analyses, and is typically based on injections campaigns into search pipelines~\cite{2016ApJ...833L...1A,2018arXiv181112940T,2018CQGra..35n5009T}. 
Here we implement a common (but accurate) approximation. We model $p_{\rm det}$ using the cumulative distribution 
of the projection parameter~\cite{1993PhRvD..47.2198F,1996PhRvD..53.2878F,2010ApJ...716..615O,2015ApJ...806..263D,2018PhRvD..98h3017T,2016ApJ...819..108B,2017arXiv170908079C} 
\begin{align}
\omega &= \sqrt{ \frac{(1+\cos^2\iota)^2}{4}F_+^2(\theta,\phi,\psi)+ \cos^2\iota F_\times^2(\theta,\phi,\psi)}\leq1\,,
\label{omega}
\end{align}
where 
\begin{align}
F_+&= \frac{1}{2}\left (1+\cos^2\theta \right )\cos 2\phi \cos 2\psi - 
\cos\theta\sin 2\phi \sin 2\psi\,, \label{Fplus}
\\
F_\times &=\frac{1}{2}\left (1+\cos^2\theta \right )\cos 2\phi \sin 2\psi + 
\cos\theta\sin 2\phi \cos 2\psi\,, \label{Fcross}
\end{align}
are the single-detector antenna pattern functions. The parameter $\omega$ is an analytic function of binary inclination $\iota$, sky location $\theta$ and $\phi$, and polarization angle $\psi$ which encapsulates all the angular dependence of the signal-to-noise ratio (SNR). The SNR  of a generic binary is given by 
$\rho=\omega\times\rho_{\rm opt}$,
where $\rho_{\rm opt}$ is the SNR of an optimally oriented source with the same parameters $\lambda$ and $z$. The probability density function $p(\omega)$ is obtained using Eqs.~(\ref{omega})-(\ref{Fcross}) and assuming that $\cos\iota$, $\cos\theta$, $\phi$ and $\psi$ are uniformly distributed. Selection effects are then implemented with a SNR threshold $\rho_{\rm  thr}$ by evaluating 
\begin{align}
p_{\rm det}(\lambda,z) = \int_{\rho_{\rm  thr}/\rho_{\rm opt}(\lambda,z)}^1  \!\!\!\!\!\!\!\!\!\! p(\omega) d\omega\,.
\label{pdetground}
\end{align}

This simplified approach, which is widely used in the literature, has been found to be a good approximation to more accurate estimates of detector selection effects based on simulated signals  and false-alarm rates (see~\cite{2016ApJ...833L...1A,2018arXiv181112940T} for comparisons). 

In the following, we compute $\rho_{\rm opt}$ using the waveform model of Ref.~\cite{2014PhRvL.113o1101H} and the noise curves of either LIGO at design sensitivity~\cite{2018LRR....21....3A} or the proposed third-generation detector Cosmic Explorer~\cite{2017CQGra..34d4001A} (hereafter ``3g'' or ``3rd gen.''). For ground-based detectors,  we set $\rho_{\rm thr}=8$~\cite{2016ApJ...833L...1A}. Equation~(\ref{pdetground}) is evaluated by Monte Carlo integration \cite{gwdet}.

\subsection{Detection rate from space}

Estimates of selection effects for space-based detectors must necessarily take into account the mission duration. According to current design choices, the nominal (extended) LISA mission duration is $T_{\rm obs}=4\,(10)$~yr~\cite{2017arXiv170200786A}.

GWs emitted by a binary that will merge in a time $t_{\rm merger}$ are detected with  frequency~\cite{1964PhRv..136.1224P}
\begin{align}
f(t_{\rm merger}) = \frac{5^{3/8}}{8\pi } [M_c (1+z)]^{-5/8} t_{\rm merger}^{-3/8}\,,
\label{mergerfreq}
\end{align}
where $M_{c}=(m_1 m_2)^{3/5}/(m_1+m_2)^{1/5}$ is the source-frame chirp mass. After a time $T_{\rm obs}$, the same source will be visible with frequency $f(t_{\rm merger} - T_{\rm obs})$, unless it has merged before. The source's SNR during the entire mission duration is given by \begin{align}
\rho^2(t_{\rm merger}) = 4 \int_{f(t_{\rm merger})}^{f[
\max(t_{\rm merger} - T_{\rm obs}),0]} \frac{|\tilde h(f)|^2}{S_n(f)} df\,.
\label{rhospace}
\end{align}
We use the sky-averaged noise curve $S_n(f)$ of Ref.~\cite{2018arXiv180301944R}. We approximate the strain $\tilde{h}$ and the merger frequency $f(0)$ as in Ref.~\cite{2007CQGra..24S.689A}. Stellar-mass BH binaries emit in the mHz regime at very wide separations, where spin  effects can be safely neglected~\cite{2018arXiv181101805M}.%

\begin{figure}[t]
\includegraphics[width=\columnwidth]{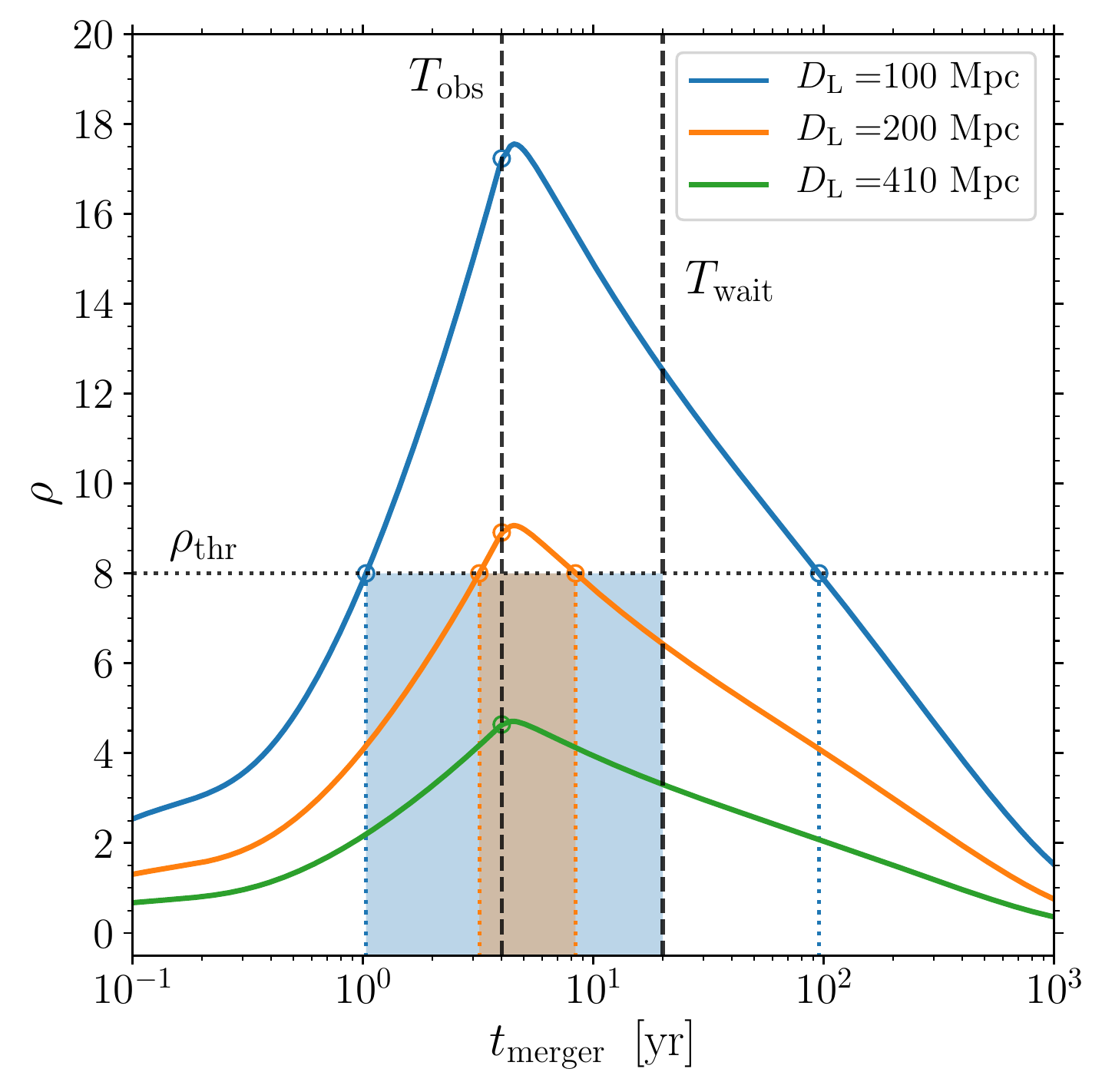}
\caption{Sky-averaged SNR $\rho$ for LISA as a function of the binary merger time $t_{\rm merger}$. Solid colored lines indicate three sources with masses of $29M_{\odot}$ and $36 M_{\odot}$, like those of GW150914~\cite{2018arXiv181112907T}, placed at luminosity distances $D_{\rm L}= 100, 200$, and $410$ Mpc (the latter being compatible with the actual GW150914 event). The optimal SNR is obtained at merger times close to the mission lifetime $T_{\rm obs}=4$ yr (dashed line). The closest sources are found above the threshold $\rho_{\rm thr}=8$ during the time window marked by the dotted lines. This can be further restricted if only sources merging within a time $T_{\rm wait}$ (dashed line) are sought for. Shaded colored regions indicate effective time window that determines the multiband rate.}
\label{timewindow}
\end{figure}

Figure~\ref{timewindow} shows the function $\rho(t_{\rm merger})$ for some representative BH binaries. The SNR is expected to decrease for large values of $t_{\rm merger}$ because the binary does not chirp much, i.e. $f(t_{\rm merger})\simeq f(t_{\rm merger} - T_{\rm obs})$. The SNR is also expected to be low if the binary is too close to merger, because $f(t_{\rm merger})$ falls outside the LISA sensitivity band. The largest SNR is obtained for $t_{\rm merger}\ssim T_{\rm obs}$, corresponding to the case where the binary spends the longest time chirping in the LISA band. As shown in Fig.~\ref{timewindow}, the maximum of $\rho$ is actually located at mergers times slightly larger than $T_{\rm obs}$, such that none of the available mission lifetime is ``wasted'' at frequencies $\gtrsim 1$~Hz where LISA is blind.
 
We now wish to impose a SNR threshold. A conservative threshold $\rho_{\rm thr}=8$ is typically considered sufficient to extract signals from the LISA data stream~\cite{1993PhRvL..70.2984C}. For the specific case of multiband observations,~\citet{2018PhRvL.121y1102W} recently pointed out that ground-based detections could be used \emph{a posteriori} to dig deeper into the LISA noise, thus lowering the effective SNR threshold to $\rho_{\rm thr}\simeq 4$. %

Let us denote with $t_{\rm thr1}$ and $t_{\rm thr,2}$ the roots (if any) of the equation
\begin{align}
\rho(t_{\rm merger}) = \rho_{\rm thr} \label{LISASNRthr}\,.
\end{align}
The quantity $|t_{\rm thr1} - t_{\rm thr2}|$ provides an estimate of the time window in which a merging BH binary is visible from space (cf. Fig.~\ref{timewindow}). The number of observations for a space-based detectors can thus be estimated by
\begin{align}
N_{\rm space}=\iint &dz d\lambda\; \mathcal{R}(z)\;p(\lambda) \frac{d V_c(z)}{dz} \frac{1}{1+z}  
\notag\\
&\;\,\times \Big|t_{\rm thr1}(\lambda, z) - t_{\rm thr2}(\lambda, z)\Big|\,,
\label{pdetspaceonly}
\end{align}
while the detection rate is 
\begin{align}
r_{\rm space}   = \frac{N_{\rm space}}{T_{\rm obs}}\,.
\label{eq:rate}
\end{align}
Let us note that $N_{\rm space}$ does not depends on $T_{\rm obs}$ explicitly, but only implicitly through the SNR.

\subsection{Multiband detection rate}

Finally, we are interested in the combined ground $+$ space detection rate. In this context, it might be desirable to impose that detections by the two instruments should happen within a time frame $T_{\rm wait}$. Longer merger times are discarded  and the effective merger time window is consequently reduced (cf. Fig.~\ref{timewindow}). The multiband detection rate is obtained by combining the ground-based detectability $p_{\rm det}$ with the space time window, restricted to merger times smaller than $T_{\rm wait}$. One obtains
\begin{align}
&N_{\rm multib} = \;\mathcal{F}   \iint dz d\lambda\; \mathcal{R}(z)\;p(\lambda) \frac{d V_c(z)}{dz} \frac{1}{1+z}\;p_{\rm det}(\lambda, z)
\notag\\
&\qquad\times \Big|\min\left[ t_{\rm thr1}(\lambda,z), T_{\rm wait}\right] - \min\left[ t_{\rm thr2}(\lambda,z), T_{\rm wait}\right]\Big|\,,
\label{pdetmultibib}
\end{align}
where $0\leq\mathcal{F}\leq 1$ is the duty cycle of the ground-based network during the space-mission lifetime.
The multiband detection rate is then given by 
\begin{align}
r_{\rm multib}  = \frac{N_{\rm multib}}{\mathcal{F}\,\,T_{\rm obs}} \,.
\label{rmultibib}
\end{align}
For simplicity, in the rest of the paper we set  $\mathcal{F}=1$ --a value which hopefully is indicative of future scenarios with ground-based networks of 4 or more instruments. The duty cycle only enters Eq.~(\ref{pdetmultibib}) linearly and cancels out in Eq.~(\ref{rmultibib}). Our results for $N_{\rm multib}$ can be trivially rescaled
to other values of $\mathcal{F}$.
Unless specified otherwise, we set $T_{\rm wait} = 5 \times T_{\rm obs}$.

\section{\mbox{$\!\!\!$Predictions based on current data}}
\label{datapredictions}

We now present some simple estimates based on the merger rate measured by LIGO/Virgo using 10 BH binary detections~\cite{2018arXiv181112907T}.  The intrinsic merger rate $\mathcal{R}$ is strongly correlated with the parameter distribution  $p(\lambda)$. In this section we only consider nonspinning BHs, i.e. $\lambda=\{m_1,m_2\}$. 

In Ref.~\cite{2018arXiv181112907T}, two mass distributions $p(m_1,m_2)$ were used.
\begin{itemize}
\item [(i)] In the first model,  primary masses are distributed according to a Salpeter power law and  secondary masses are distributed uniformly: $p(m_1)\propto m_1^{-2.3}$,
$p(m_2| m_1)={\rm const}$. The merger rate was measured to be $\mathcal{R}= 57^{+40}_{-25}$ Gpc${^{-3}}$yr${^{-1}}$~\cite{2018arXiv181112940T} (where  errors refer to 90\% credibility and results from two independent pipelines have been combined).
\item [(ii)]  In the second model, both masses are distributed uniformly in log: $p(m_1,m_2)\propto 1/m_1 m_2$. This choice lowers the measured  rate to  $\mathcal{R}= 19^{+13}_{-8.2}$ Gpc${^{-3}}$yr${^{-1}}$~\cite{2018arXiv181112940T}.
\end{itemize} 
We refer to these variations as ``powerlaw'' and ``log'', respectively. 
For each of these two choices, we take the median\footnote{Fig.~12 in Ref.~\cite{2018arXiv181112907T} suggests that the posterior distribution of $\mathcal{R}$ is approximately log-normal. An estimate of the expectation values $\int \mathcal{R} d  \mathcal{R}$  based on the reported 90\% confidence interval returns values within $5\%$ of the medians.} as well as the lower and upper edges of the 90\% confidence interval. This results in three estimates for $\mathcal{R}$ that we refer to as ``median'', ``lower'' and ``upper''.  We assume $m_1,m_2\in [5 M_\odot,50 M_\odot]$ as in Ref.~\cite{2018arXiv181112907T}.

We integrate Eq.~(\ref{rground}) with standard Monte-Carlo methods to obtain detection rates $r$ and number of observations $N$  for ground-based detectors, space missions and multiband scenarios. %
Our results are summarized Fig.~\ref{overviewNobs}. 
Additional details are reported in Appendix~\ref{suppmat}.

\begin{figure}
\includegraphics[width=\columnwidth]{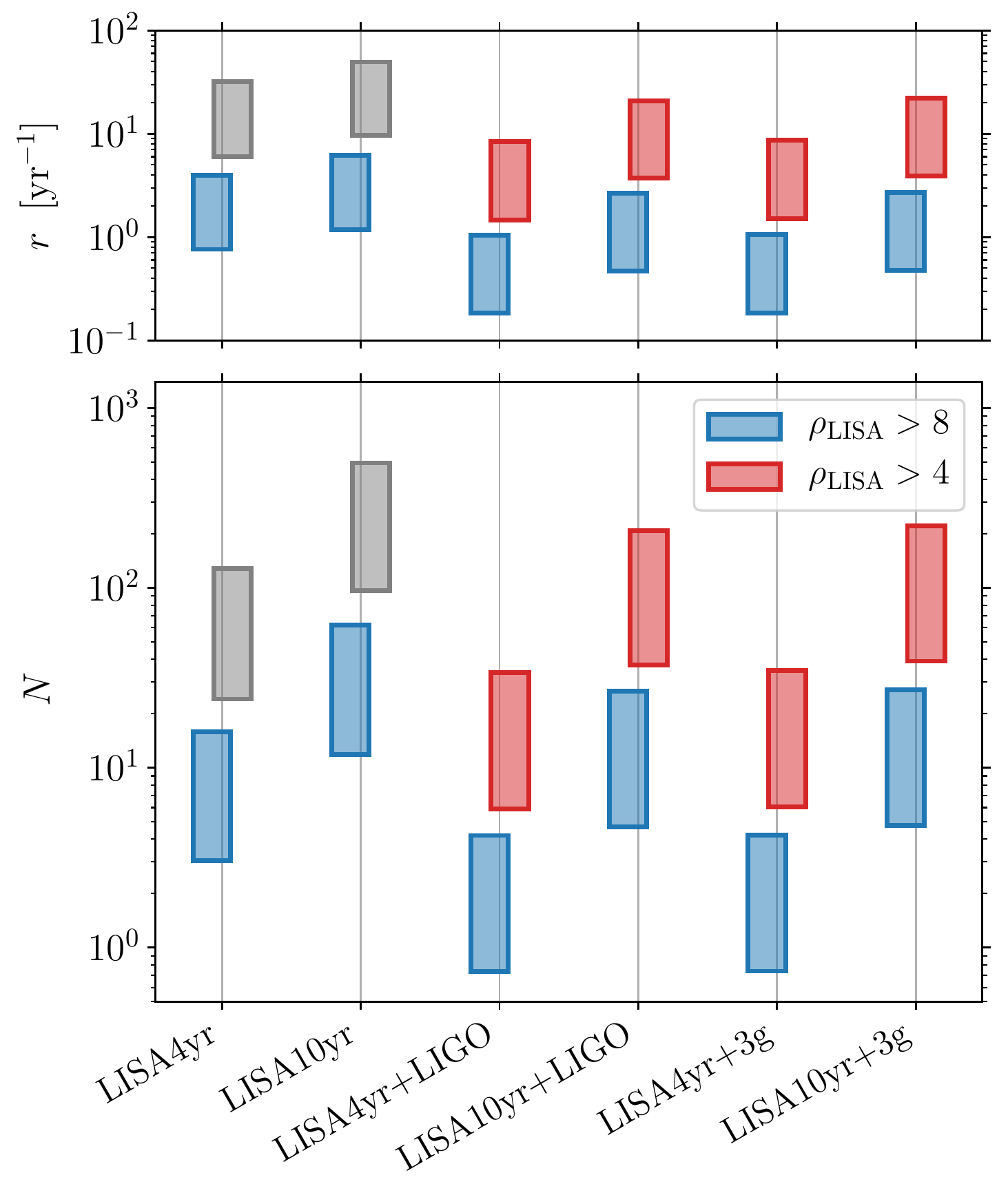}
\caption{Expected detection rates (top) and number of observations (bottom) of stellar-mass BH binaries observable from either space-based detectors alone or space- and  ground-based observatories. We consider two possible outcomes of the LISA mission with duration of $4$ and $10$ yr, together with current (LIGO) and future (Cosmic Explorer, ``3g'') ground-based instruments. We present results obtained by extremizing over two mass spectra (``powerlaw'' and ``log'') and three estimate of the intrinsic merger rate (``lower'', ``median'', ``upper''); see text for details. We assume two possible LISA SNR  thresholds, meant to indicate scenarios where multiband detections are realized first from space and then from the ground ($\rho_{\rm LISA}>8$, blue), or vice versa ($\rho_{\rm LISA}>4$, red). The latter case is not possible for  LISA alone unless data analysis techniques dramatically improve (gray).}
\label{overviewNobs}
\end{figure}

We estimate that LISA will observe $3-12$ stellar-mass BH binaries with SNR greater than 8 during its nominal mission of duration $T_{\rm obs}=4$ yr. An extension to $T_{\rm obs}=10$ yr will deliver $10-50$ sources in total. As for multiband detections, a 4 yr (10 yr) LISA mission will provide forewarnings to ground-based observatories for a number of events between 0 and 4 (4 and 22). Note that, in this case, we only consider binaries that merge within a time $T_{\rm wait} = 5 \times T_{\rm obs}$ [cf. Eq.~(\ref{pdetmultibib})]

Weaker signals could be extracted from the LISA data stream by leveraging the information provided by ground-based interferometers~\cite{2018PhRvL.121y1102W}. With a lower threshold $\rho_{\rm LISA}>4$, the number of multiband events increases by about a factor of $\ssim 5$. We predict $N\ssim 5-30$~~($35-170$) for $T_{\rm obs}=4$ yr (10 yr). While the number of detections for which LISA will be able to predict the merger time (i.e. $\rho_{\rm LISA}>8$) is compatible with zero in the most pessimistic cases, our analysis confidently predict multiple binaries with $\rho_{\rm LISA}>4$ that can be extracted \emph{after}  ground-based detections have been made.

Interestingly, the noise level of the ground-based instrument does not play a role in this estimate. A ``LISA+LIGO'' or a ``LISA+3g'' network will deliver the same multiband binaries. This feature can be understood in terms of the horizon redshifts of the instruments involved (cf. Appendix \ref{zhorsec}). LIGO (3g) can observe a $30+30 M_\odot$ binary up to $z\ssim 1.2$ ($35$), while LISA is restricted to $z\lesssim 0.4$. All the sources that are invisible to LIGO but will be observed by a third-generation detector are invisible to LISA as well. Obviously, more sensitive detectors will allow for more accurate characterizations of the sources. But regarding multiband detections rates, second- and third-generation detectors behave essentially the same.

\begin{figure}[t]\centering
\includegraphics[width=\columnwidth]{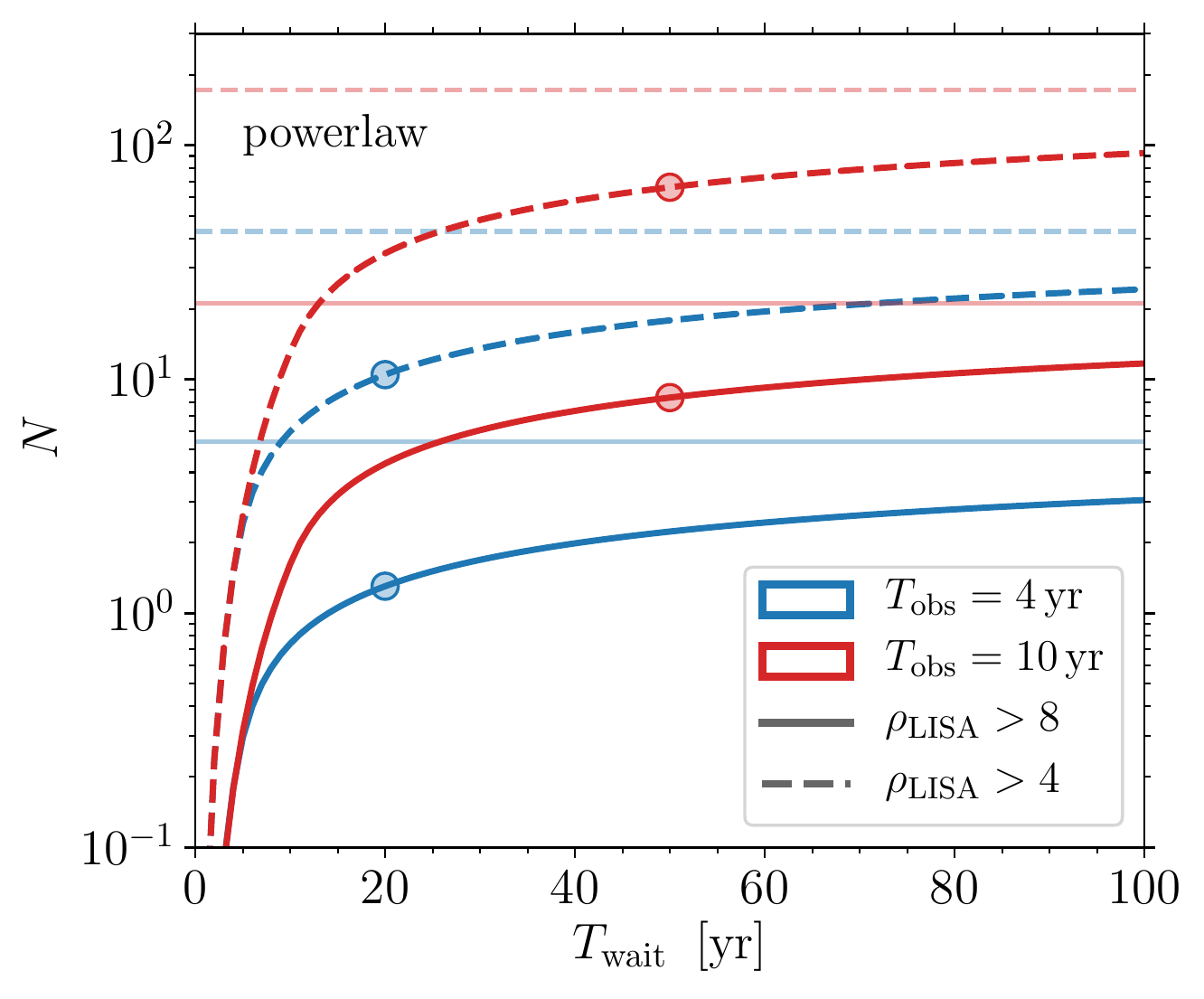}
\caption{Number of multiband events as a function of the time lag between space-and ground-based detection. We assume a LISA mission duration of $T_{\rm obs}=4$~yr (blue) and $10$~yr (red) and a  LISA SNR threshold of 8 (solid) and 4 (dashed). Circles mark our default choice $T_{\rm wait}=5\times T_{\rm obs}$. Shaded horizontal lines indicate the number of events predicted by the LISA mission alone, irrespectively of whether they are observable from the ground. This figure is produced assuming the ``powerlaw'' mass spectrum and the ``median'' assumption for the intrinsic merger rate. Results are very similar if LIGO or a third-generation ground-based detector is considered; for concreteness, here we use LIGO. }
\label{NTwait}
\end{figure}

The time lag between ground- and space-based observations %
 is investigated in Fig.~\ref{NTwait}, where we show the number of expected events as a function of $T_{\rm wait}$.  For very large waiting times, the number of multiband observations $N_{\rm multib}$ approaches the value predicted for the LISA mission alone $N_{\rm space}$. However, it does not exactly tend to it because of selection effects in the ground-based detector: Eq.~(\ref{pdetspaceonly}) and Eq.~(\ref{pdetmultibib}) differ by a factor $p_{\rm det}(\lambda,z)$ as $T_{\rm wait}\to \infty$. As outlined above, the sensitivity of the ground-based detector hardly matters  and the difference between $N_{\rm space}$ and $N_{\rm multib}(T_{\rm wait} \to \infty)$ is consequently very small. Figure~\ref{NTwait} shows that one only needs to wait for a time comparable to the mission lifetime to observe a good fraction of the multiband binaries. 
 The number of observations rises steeply for $T_{\rm wait}\lesssim T_{\rm obs}$, and then flattens for larger values (our default choice $T_{\rm wait}=5\times T_{\rm obs}$ falls in the latter regime).  The role of $T_{\rm wait}$ is to filter out binaries with large merging times (cf. Fig.~\ref{timewindow}). Those same binaries do not chirp much and are, therefore, likely to have SNRs below threshold. 
 In particular, we find $N_{\rm multib}(T_{\rm wait} = T_{\rm obs})/N_{\rm space}\simeq 10\%$.

\begin{figure*}[t]\centering
\includegraphics[width=\textwidth]{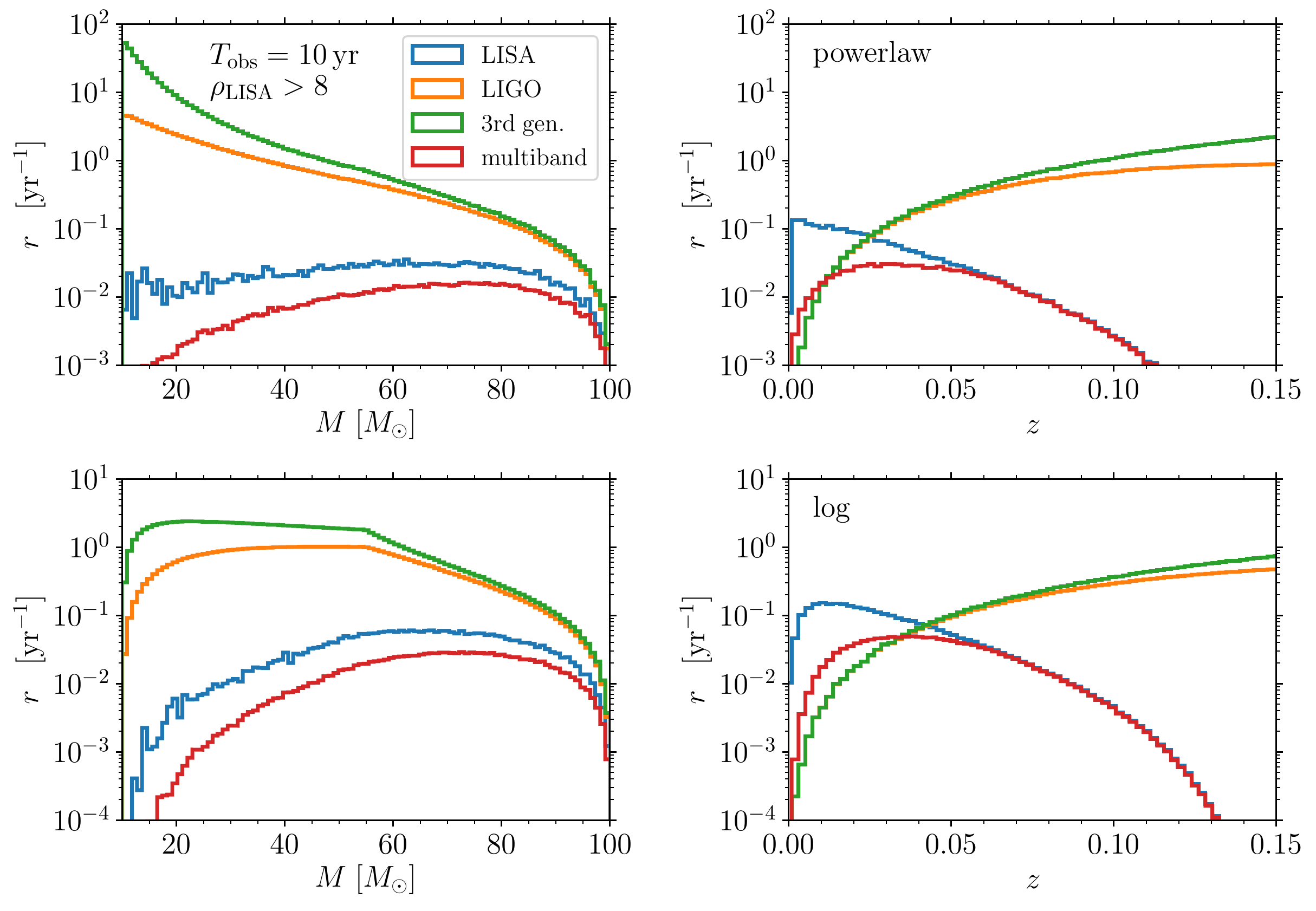}
\caption{Marginalized event-rate distribution of total mass $M=m_1+m_2$ (left panels) and redshift $z$ (right panel) of stellar-mass BH binaries detectable by LISA, LIGO, Cosmic Explorer (``3rd gen.''), and multiband scenarios. 
On the y-axis we show $dr/dM$ and $dr/dz$ from Eq.~(\ref{rground}) times the bin sizes $\Delta M$ and $\Delta z$, such that the sum of the histogram entries is equal to the total rate $r$. Multiband distributions are identical if a ``LISA+LIGO'' or a ``LISA+3g'' network is considered. Results are restricted to the local Universe ($z<0.3$) and obtained using the ``median'' value of $\mathcal{R}$ from Ref.~\cite{2018arXiv181112907T}. We assume the ``powerlaw'' (top) and ``log'' (bottom) mass spectra, a LISA mission duration 10 yr, and a SNR threshold of 8 (see text for details).}
\label{dratedMz}
\end{figure*}

Figure~\ref{dratedMz} shows the detectable distributions of total masses $M$ and  redshifts $z$ predicted in these scenarios. For concreteness, we use a LISA SNR threshold of 8, but results hold qualitatively for $\rho_{\rm LISA}>4$ as well.
 A few interesting trends are present.
First, LISA observations strongly select binaries with total mass $M\gtrsim 60M_\odot$. The GW emission frequency $f$ is related to the total mass and the binary separation $r$ by $f=\sqrt{M/\pi^2 r^3}$: low-mass sources emit in the LISA band when their orbital separation is too large for their GW radiation to be detectable.
Second, all stellar-mass BH binaries detectable by LISA are relatively local, with redshifts $z\lesssim 0.1$ (cf. Appendix \ref{zhorsec}). 
Among all the stellar-mass BH binaries detectable from the ground, only those with  $M\gtrsim 60 M_\odot$ and $z\lesssim 0.1$ are accessible from space.

The case of multiband detections is not only determined by the LISA sensitivity, but also by the time between the two sets of observations one is willing to wait. In this paper, we denote as ``multiband'' only those binaries for which space and ground observations are  separated at most by a time $T_{\rm wait}$ (which is set to $50$ yr in Fig.~\ref{dratedMz}). The multiband detection rate is lower than the rate for LISA alone  because binaries with merger time longer than  $T_{\rm wait}$ are discarded. 

This feature is particularly evident in the $dr/dz$ distribution (right panels of Fig.~\ref{dratedMz}). For  $z\lesssim 0.02$, the LISA detection rate exceeds that of ground-based detectors. This might seem counterintuitive at first, but it is actually expected because binaries spend a longer time at the low frequencies accessible from space [$t\propto f^{-8/3}$ from Eq.(\ref{mergerfreq})]. For sources at very low redshifts, the effective merger time window  $|t_{\rm thr1} - t_{\rm thr2}|$ of Eq.~(\ref{pdetspaceonly}) can be as long as $\ssim 10^4$ yr. The threshold $T_{\rm wait}$ largely removes this effect. If binaries with very long merger time are considered not interesting, the predicted multiband detection rate approaches those of ground-based detectors at low redshifts.
As stressed above, the sensitivity of the ground based interferometer plays a minor role: in other terms, $p_{\rm det}\lesssim 1$ whenever Eq.~(\ref{LISASNRthr}) admits roots.

\section{Population synthesis predictions}
\label{spopssection}

We now apply our rate analysis to state-of-the-art population synthesis simulations of spinning BH binaries formed from binary stars. In particular, we only consider isolated systems that evolved through a common-envelope phase (for a review on BH binary formation channels see, e.g., Ref.~\cite{2018arXiv180605820M}). We use publicly available distributions presented in Refs.~\cite{2018PhRvD..98h4036G,spops}.  We perform stellar evolutions with the \textsc{Startrack}~\cite{2008ApJS..174..223B,2017arXiv170607053B} population-synthesis code, and add spins in postprocessing using the prescriptions of Refs.~\cite{2013PhRvD..87j4028G,2018PhRvD..98h4036G} and the \textsc{precession}~\cite{2016PhRvD..93l4066G} code. %

Among the spin variations presented in Ref.~\cite{2018PhRvD..98h4036G}, for simplicity we restrict ourselves to the ``time-uniform'' model, i.e. we adopt a physically motivated prescription for tidal alignment timescale and we assume that BH spins are uniformly distributed in magnitude between zero and the Kerr bound. We vary a single population parameter, namely the magnitude of the kicks imparted at BH formation~\cite{1999ApJ...526..152F}. Natal kicks are distributed according to a Maxwellian distribution with 1D dispersion $\sigma\in [0,265]$ km/s, independently of the star's mass. This is a simple, one-parameter family of models that allows us to bracket the expected detection rates~\cite{2002ApJ...572..407B,2018PhRvD..98h4036G}.

Our dataset contains distributions of BH masses and spins, i.e. $\lambda = \left\{m_1,m_2, \boldsymbol{\chi}_1,\boldsymbol{\chi}_2 \right\}$. We integrate Eq.~(\ref{rground}) with Monte Carlo techniques accounting for the expected distributions of zero-age-main-sequence stars, as well as the expected redshift- and metallicity-dependent star-formation rate~\cite{2016Natur.534..512B,2016ApJ...819..108B}.

\subsection{Multiband rates and number of detections}

\begin{figure} \includegraphics[width=\columnwidth]{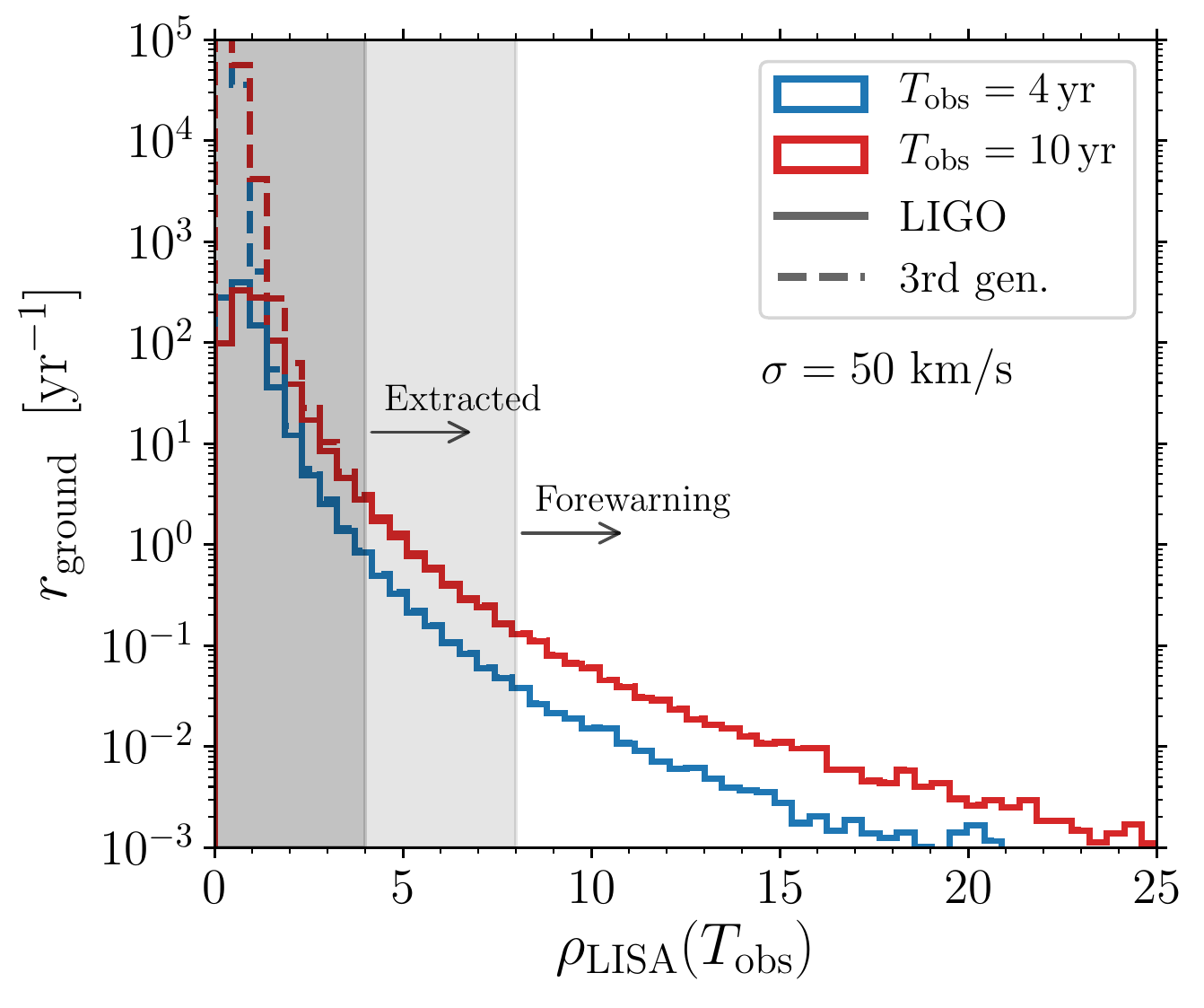}
\caption{Distribution %
BH binaries detectable from the ground as a function of the SNR in LISA. The x-axis shows $\rho_{\rm LISA}$ evaluated at $T_{\rm obs}$ (as a proxy for the strength of the signal), while the y-axis shows the ground-based detection rate per bin.  Results are presented for a single population synthesis simulation with supernova kick parameter $\sigma=50\;{\rm km/s}$. We assume a LISA mission duration of $T_{\rm obs}=4$~yr (blue) and $10$~yr (red), and compute ground-based selection effects using either LIGO (solid) or Cosmic Explorer (dashed).}
\label{SNRdistr}
\end{figure}

\begin{figure}
\includegraphics[width=\columnwidth]{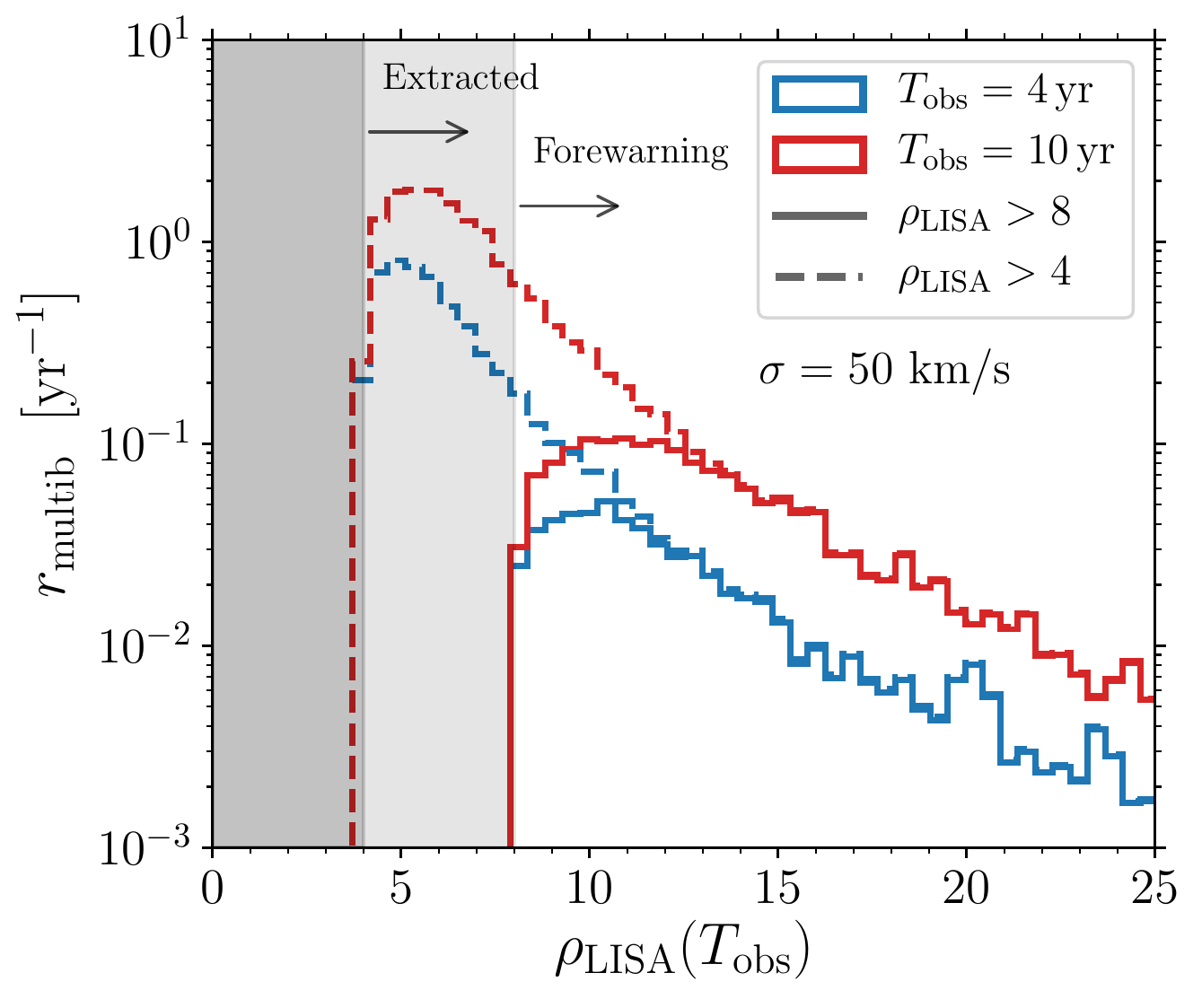}$\quad$
\caption{Distribution of LISA SNRs for multiband BH binaries. The x-axis shows $\rho_{\rm LISA}$ evaluated at $T_{\rm obs}$ (as a proxy for the strength of the signal), while the y-axis shows the multiband detection rate per bin. Results are presented for a single population synthesis simulation with supernova kick parameter $\sigma=50\;{\rm km/s}$. We assume a LISA mission duration of $T_{\rm obs}=4$~yr (blue) and $10$~yr (red) and a LISA SNR threshold of 8 (``forewarning'', solid) and 4 (``extracted'', dashed). Results appear identical if LIGO or a third-generation ground-based detector is considered. We set $T_{\rm wait} = 5 \times T_{\rm obs}$, i.e. either 20 or 50 yr.}
\label{SNRthr}
\end{figure}

\begin{figure}
\includegraphics[width=\columnwidth]{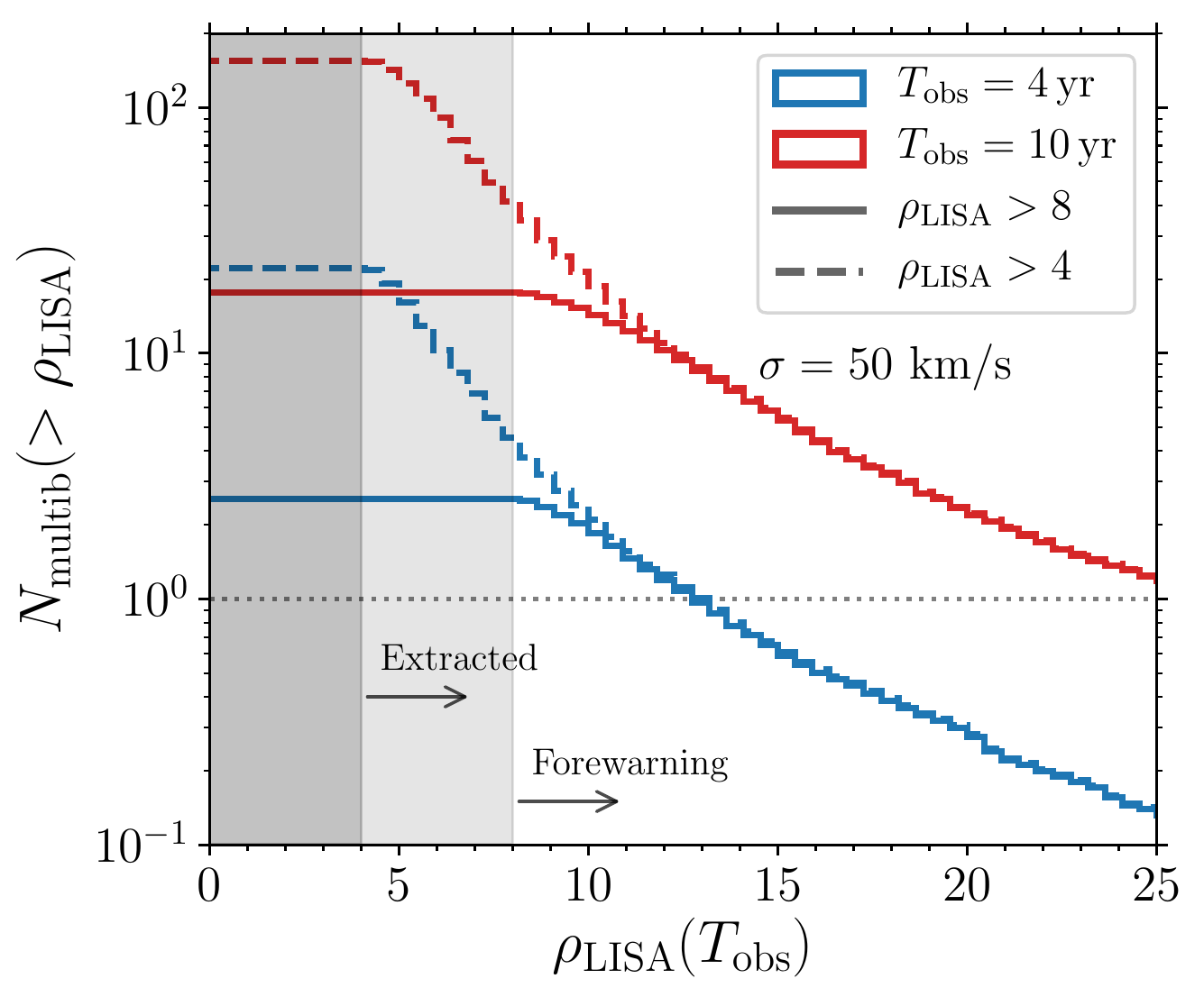}
\caption{
Cumulative number of expected multiband detections as a function of the LISA SNR.
Results are presented for a single population synthesis simulation with supernova kick parameter $\sigma=50\;{\rm km/s}$. We assume a LISA mission duration of $T_{\rm obs}=4$~yr (blue) and $10$~yr (red) and a  LISA SNR threshold of 8 (``forewarning'', solid) and 4 (``extracted'', dashed). Results appear identical if LIGO or a third-generation ground-based detector is considered. We set $T_{\rm wait} = 5 \times T_{\rm obs}$, i.e. either 20 or 50 yr. }
\label{Neventscum}
\end{figure}

Figures~\ref{SNRdistr}, \ref{SNRthr} and \ref{Neventscum} show  results for a single representative stellar evolution model with $\sigma=50\;{\rm km/s}$. This choice yields detection rates consistent with LIGO observations, and it is in agreement with current mass and spin misalignment constraints~\cite{2017PhRvL.119a1101O,2017arXiv170607053B,2018PhRvD..97d3014W} (see also Refs.~\cite{2015MNRAS.453.3341R,2016ApJ...819..108B,2016MNRAS.456..578M} for kick measurements with x-ray binaries data). We present rates and numbers of detected binaries as a function of $\rho_{\rm LISA}$ evaluated at $t_{\rm merger} =T_{\rm obs}$, cf.  Eq.~(\ref{rhospace}). This value can be considered as a simple estimator of the strength of the LISA signal, but it is understood that detection rates are computed by varying over $t_{\rm merger}$, as detailed in Sec.~\ref{methodrates}.

In Fig.~\ref{SNRdistr} we illustrates the expected LISA SNR distribution of GW sources detectable from the ground (i.e., histogram entries are weighted by the ground-based detection rate $r_{\rm ground}$). Crucially, we find that the rate is a very steep function of $\rho_{\rm LISA}$.  Lowering the SNR threshold from $\ssim 8$ to $\ssim 4$ raises the ground-based rate by a factor $\ssim (8/4)^3=8$.
Any data-analysis or experimental technique which increases, even marginally, the LISA redshift horizon is expected to have significant consequences for multiband detections.  According to Eq.~\eqref{eq:rate}, extending the mission duration from 4 to 10 yrs will increase the accessible rate by a factor of $\ssim 3$, thus delivering $\ssim 3 \times 10/4= 7.5$ times more detected sources.

Compared to LIGO, the increased sensitivity of third-generation detectors is only relevant for binaries with $\rho_{\rm LISA}\lesssim 2$, which are too faint to be identified in the LISA data. For concreteness, in the following we show multiband rates computed using the LIGO noise curve, but results would be unchanged if we considered Cosmic Explorer instead. 

Figure~\ref{SNRthr} shows the LISA SNR distribution restricted to those fewer (multiband) sources that are accessible from both ground and space. As expected, the multiband rate steeply decreases as the SNR threshold  is approached. It is worth noting, however, that the distributions of Fig.~\ref{SNRthr} do not present a sharp and unphysical cutoff, but they decrease smoothly as $\rho\to \rho_{\rm thr}$. This
is a direct consequence of the procedure presented in Sec.~\ref{methodrates}: LISA observes for a time $T_{\rm obs}$ but, in general, sources will not inspiral all the way to merger during that time. Their effective SNR is, therefore, somewhat smaller than $\rho_{\rm LISA}(T_{\rm obs})$.

A summary of these results is presented in Fig.~\ref{Neventscum}, where we show the cumulative distribution of the number of multiband events. For this specific model with $\sigma=50$~km/s, a LISA mission of $T_{\rm obs}=4$ (10) yr will deliver $\ssim 2$ ($\ssim 20$) multiband sources with SNR greater than 8. About $\ssim 20$ ($\ssim 130$) more sources will be seen with $4<\rho_{\rm LISA}(T_{\rm obs})<8$.

\begin{figure}
\includegraphics[width=\columnwidth]{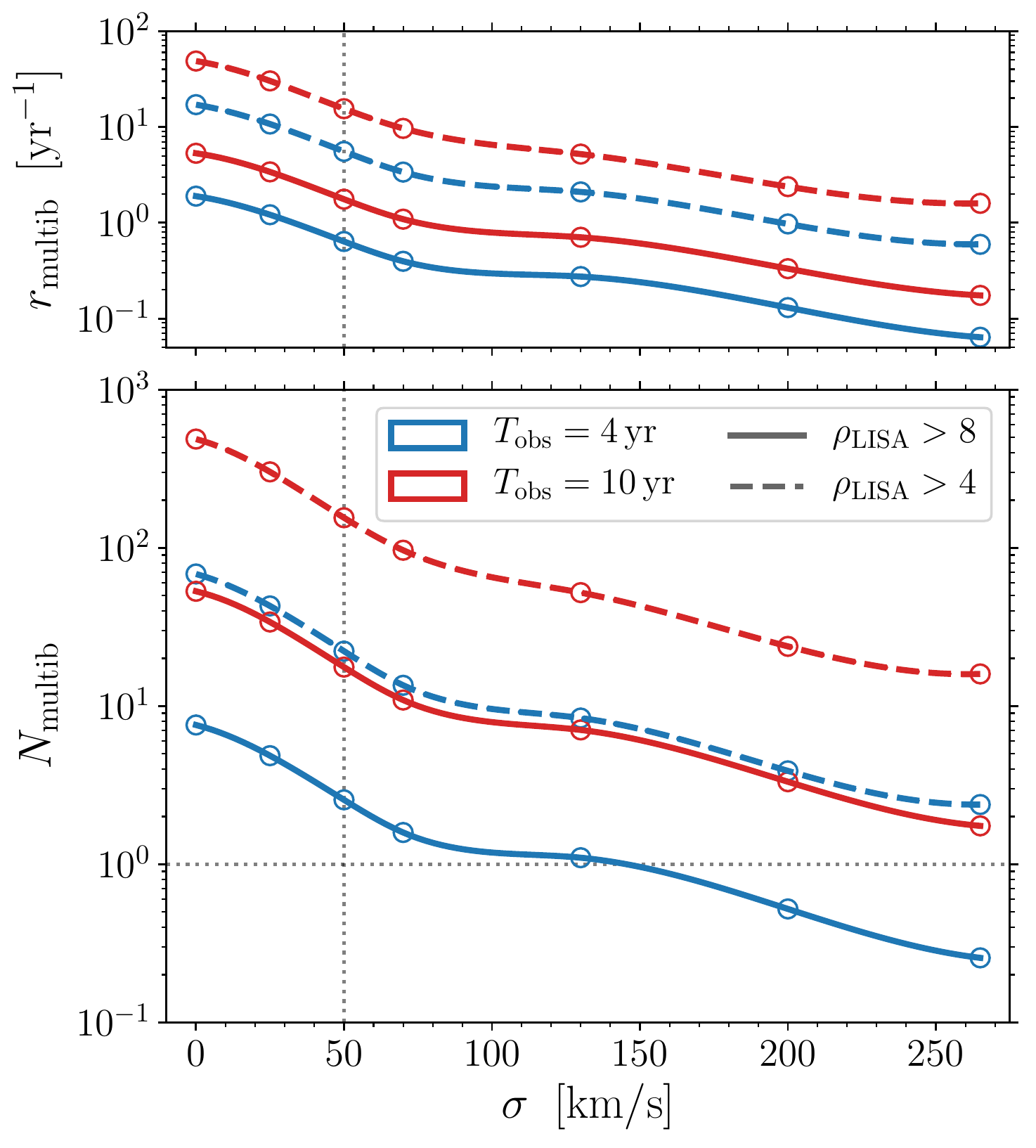}
\caption{Multiband detection rates (top) and number of events (bottom) predicted by population synthesis simulations of binary stars. We present results from 7 different simulations with varying supernova kicks, where the 1D dispersion $\sigma$ ranges from $0$~km/s (no kicks) to $265$~km/s (BHs receives full kicks as inferred from galactic pulsars). Blue (red) curves show results assuming a LISA mission duration of 4~yr (10~yr); solid (dashed) lines assumes a LISA SNR threshold of 8 (4). The vertical dotted line at $\sigma=50$~km/s marks the model used in Figs.~\ref{SNRdistr}-\ref{Neventscum}. Results appear identical if LIGO or a third-generation ground-based detector is considered. We set $T_{\rm wait} = 5 \times T_{\rm obs}$, i.e. either 20 or 50 yr.  Data to reproduce this figure are reported in Table~\ref{startracktable}.}
\label{foreextkick}
\end{figure}

We explore the uncertainties of our predictions on $r_{\rm multib}$ and $N_{\rm multib}$ in Fig.~\ref{foreextkick}, where we vary the strength of the kicks imparted at BH formation from $\sigma=0$~km/s to $265$~km/s~\cite{2005MNRAS.360..974H} (see Appendix~\ref{suppmat} for additional details). As $\sigma$ increases, more and more binaries are unbound by supernova explosions, and the detection rate consequently decreases. We report variations of a factor $\ssim 30$. It is worth noting that only moderately extreme cases ($\sigma\gtrsim 150$~km/s, $T_{\rm obs}=4$~yr) present $N_{\rm multib}<1$. Therefore, we 
predict that the LISA space mission will detect a few -- but possibly dozens -- sources which are also observable from the ground. These events (with $\rho_{\rm LISA}>8$) are going provide forewarnings for ground-based operations, but about $\mathcal{O}(100)$ more events with $\rho_{\rm LISA}>4$ could potentially be extracted from the LISA data stream in coincidence with sources that have already been observed from Earth.

\subsection{Stellar progenitors}

Multiband GW detections offer new opportunities to explore the physics of massive stars (see e.g. Refs.~\cite{2016PhRvD..94f4020N,2016ApJ...830L..18B,2018arXiv180208654S}). By leveraging our \textsc{Startrack}+\textsc{precession} simulations~\cite{2018PhRvD..98h4036G}, we now investigate which specific formation subchannels are more likely to produce multiband sources.

\begin{figure*}\centering
\includegraphics[width=0.8\textwidth]{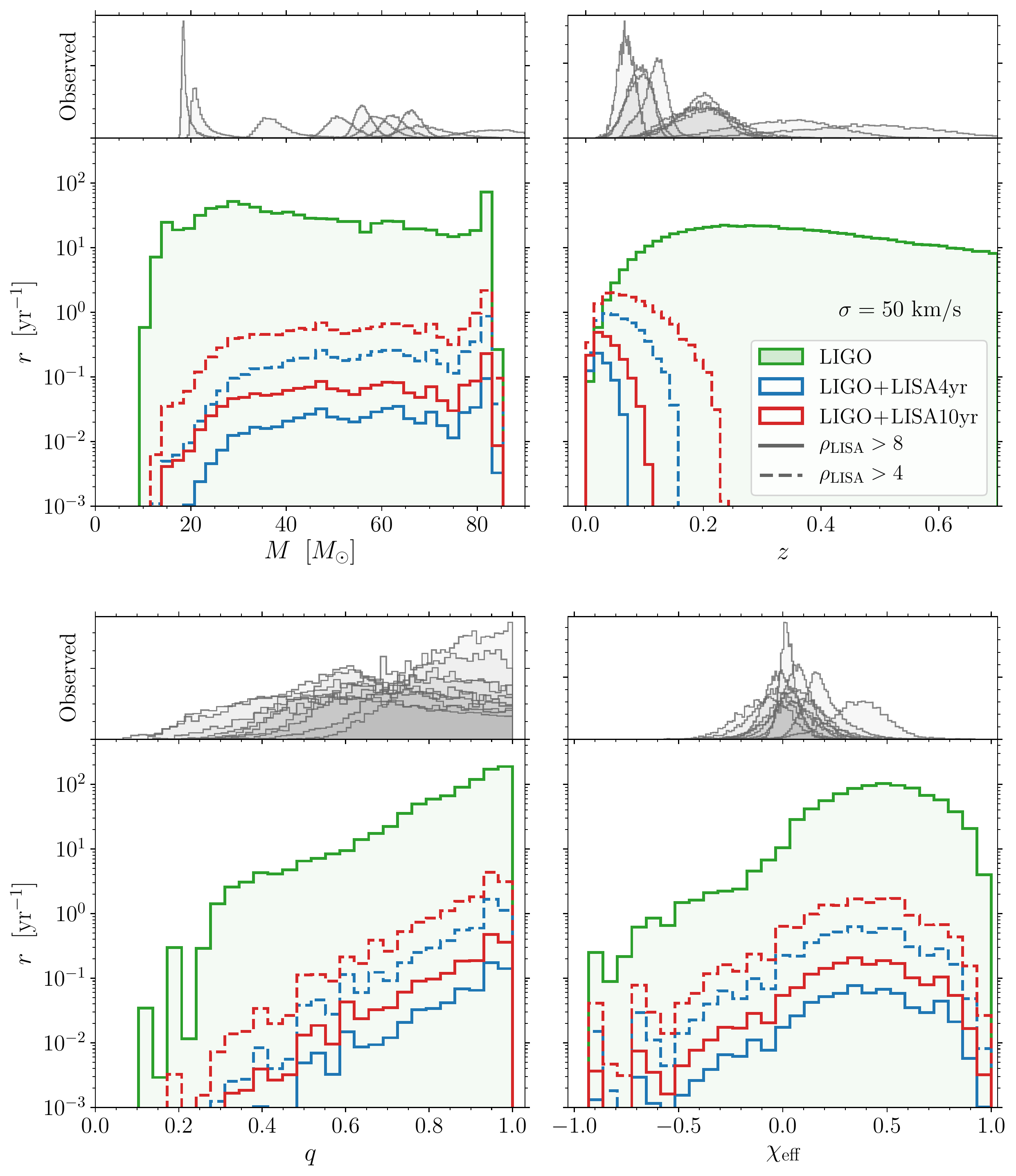}
\caption{Marginalized distributions of total mass $M=m_1+m_2$ (top left), mass ratio $q=m_2/m_1$ (bottom left), redshift $z$ (top right) and effective spin $\chi_{\rm eff}$ (bottom left) detectable by multiband GW networks. Larger panels show results from a population synthesis distribution with supernova kick parameter $\sigma=50$ km/s. The LIGO detection rate is shown in green, while multiband rates are shown in blue and red assuming a LISA mission duration of 4 and 10 years, respectively. For the multiband distributions, solid (dashed) lines are obtained with a LISA SNR threshold of 8 (4); results appear identical if LIGO or a third-generation ground-based detector is considered. The top smaller panels shows posterior distributions of the first 10 binary BH mergers detected by LIGO/Virgo~\cite{2018arXiv181112907T}. 
}
\label{massfilter}
\end{figure*}

Figure~\ref{massfilter} shows the detectable distributions of total mass, mass ratio, redshift and effective spin~\cite{2008PhRvD..78d4021R,2015PhRvD..92f4016G,2018arXiv181112907T} for our simulation with $\sigma={\rm 50}$ km/s. Binaries are weighted by detectability with either ground-based detectors alone, or multiband networks. The main trends are driven by the same considerations already presented in Sec.~\ref{datapredictions}: among the ground-based sources, only those with large enough mass and small enough redshift are detectable by LISA and can be multiband sources. 

The results of Fig.~\ref{massfilter} let us immediately infer the value of the multiband horizon redshift averaged over a suitable stellar population. For $T_{\rm obs}=4$ (10) yr, we find that only binaries at $z\lesssim 0.07$ (0.12) can be observed with $\rho_{\rm LISA}>8$ and provide forewarnings for ground-based operations. The largest redshift increases by about a factor 2 if the SNR threshold is relaxed to $\rho_{\rm LISA}>4$. These findings are in agreement with the horizon redshifts of the LISA detector alone (Appendix~\ref{zhorsec}), thus confirming that the space-based detector sensitivity is going to be limiting component of future multiband networks.

The high-mass filter is also a rather strong effect. For $M\lesssim 30 M_{\odot}$, the multiband detection rate is 3-5 orders of magnitude smaller than the LIGO one.  While only $\ssim 28\%$ of the LIGO detection rate comes from binaries with $M>60M_\odot$, that fraction increases to $\ssim 50\%$ for multiband observations. 
Multiband detections do not appear to prefer specific values of mass ratios and/or spins (Fig.~\ref{massfilter}). This is expected, because both of these parameters enter the waveform at relatively high post-Newtonian order, and the effects of high post-Newtonian orders are negligible at low frequencies.

The key stellar mechanism to form GW sources from massive binary stars in galactic fields is the so-called \emph{common-envelope} phase~\cite{1976IAUS...73...75P,2013A&ARv..21...59I}. As one of the two stars enters its late evolutionary stages, its outer layers can engulf the binary's orbit. Two stellar cores evolve inside a single envelope, which dissipates its binding energy by shrinking the binary separation.

As in Ref.~\cite{2018PhRvD..98h4036G}, we divide the results of our simulations into eight mutually exclusive subchannels depending on the formation time of the more (BH1) and less (BH2) massive objects, as well as the occurrence of the common-envelope phase (CE). The more massive BH usually originates from the more massive star and form first (i.e. ``BH1'' comes before ``BH2''). For a minority of the systems, mass transfer might reverse the binary mass ratio, causing the secondary BH to form first (i.e. ``BH2''  before ``BH1'').  Common envelope typically occurs either in between the two supernovae  (i.e. ``BH CE BH''), or  before the first one (i.e. ``CE BH BH'').  The relative importance of these two subchannels crucially depends on the magnitude of the supernova kicks~\cite{2018PhRvD..98h4036G}. For $\sigma \lesssim 100$ km/s, the dominant channel is ``BH1 CE BH2'', with the common-envelope phase involving the inspiral of an already formed BH and the core of the other progenitor. For $\sigma \gtrsim 100$ km/s, the first supernova kick is likely to unbind the stellar binary,  unless it has hardened before. In this case, the majority of the detection rate comes from the ``CE BH1 BH2'' subchannel. Other subchannels with zero or two common-envelope phases are, in general, highly subdominant.

\begin{figure*}\centering
\renewcommand{\arraystretch}{3}
\begin{tabular}{r@{\hskip 10pt}r}
\includegraphics[width=0.5\textwidth,page=1]{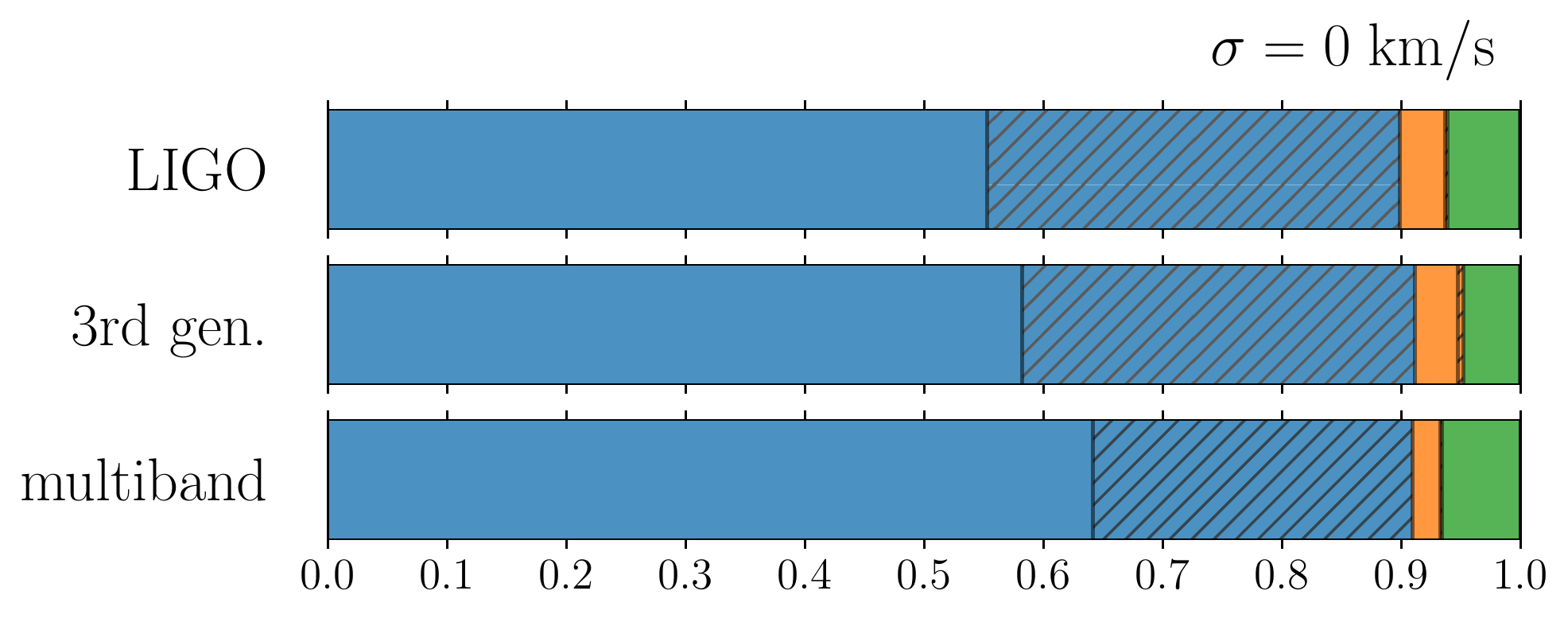}
&
\includegraphics[width=0.5\textwidth,page=2]{channelbars}
\\
\includegraphics[width=0.5\textwidth,page=3]{channelbars}
&
\includegraphics[width=0.5\textwidth,page=4]{channelbars}
\\
\includegraphics[width=0.5\textwidth,page=5]{channelbars}
&
\includegraphics[width=0.5\textwidth,page=6]{channelbars}
\\
\includegraphics[width=0.5\textwidth,page=7]{channelbars}
&
\includegraphics[trim={0 -0.8cm 0.5cm 0},width=0.41\textwidth,page=8]{channelbars}
\end{tabular}
\renewcommand{\arraystretch}{1.2}

\caption{Fractional contribution to the detection rate from BH binaries formed via different subchannels. Each panel contains results from a different population synthesis simulation, where the strength of the supernova kicks is varied from $\sigma=0$ (top left) to $265$ km/s (bottom). Binaries are classified according to the order of the formation of the heavier object (BH1), the formation of the lighter object (BH2) and the occurrence of a common-envelope phase (CE).  Top, middle and bottom bars of each panel refer to LIGO, Cosmic Explorer and  multiband detections, respectively. Multiband results are computed assuming $T_{\rm obs}=10$ yr, $\rho_{\rm LISA}>8$ and the LIGO ground-based detector (but result are unchanged if a LISA+3g network is considered).}
\label{channelbars}
\end{figure*}

Figure~\ref{channelbars} shows how the fraction of detectable sources in each of these pathways is impacted by multiband detections.
Scenarios with early common envelopes (``CE BH BH'') are generally overrepresented in multiband observations compared to
either LIGO or third-generation detectors alone.  These sources are, on average, more massive than their ``BH CE BH''
counterparts, and therefore more likely to be detected by LISA.  The ``CE BH BH'' scenarios include the most massive
stellar progenitors and BHs formed at  low metallicity, while the ``BH CE BH'' binaries form copiously at all metallicities
from pairs of $O(40 M_\odot)$ stars.  The sensitivity of third-generation detectors peaks at lower masses compared to LIGO, and it will allow them to observe more ``BH CE BH'' binaries.

Sources where BH1 forms before BH2 are not only more numerous, but also heavier than those where the lighter object forms first. Strong mass transfer is required to reverse the binary mass ratio and form the lighter BH from the heavier star. Since mass transfer is expected to be nonconservative, this implies that more mass is, on average, lost in the ``BH2 BH1'' scenarios \cite{2016Natur.534..512B}.  As shown in Fig.~\ref{channelbars}, those subchannels are slightly more probable in multiband detections compared to ground-based observations. 

\pagebreak
The few binaries that manage to merge without a common-envelope phase (i.e. ``BH BH'') also appear to be somewhat preferred by multiband scenarios. The statistics in the simulations of Ref.~\cite{2018PhRvD..98h4036G}, however, is too small to draw confident conclusions from this observation.

To summarize, the main trends observed in Fig.~\ref{channelbars} can all be understood in terms of the source total mass and its impact on LISA detectability. For reference, in our model with $\sigma=50$~km/s the medians of the total mass $M$ weighted by LIGO detections rates are $\ssim44 M_\odot$ for ``BH1 CE BH2'', 
$\ssim38 M_\odot$ for ``BH2 CE BH1'',
$\ssim54 M_\odot$ for ``CE BH1 BH2'', and 
$\ssim49 M_\odot$ for ``CE BH2 BH1''.

\section{Comparisons and prospects}
\label{conclusions}

We presented a new, semianalytic approach to compute detection rates for multiband GW sources by properly combining selection effects of ground- and space-based detectors. Our treatment relies on (i) labeling sources by their merger time, and (ii) defining an ``effective time window'' where their SNR is above threshold [see Eq.~(\ref{pdetspaceonly})]. We first exemplified the method by using the relatively model-independent LIGO/Virgo estimate of the intrinsic merger rate, and then we explored in detail the population of multiband GW sources predicted by simulations of isolated binary stars. We plan to extend our analysis to dynamical formation scenarios in future work.

There were some previous attempts at estimating multiband detection rates~\cite{2016PhRvL.116w1102S,2017JPhCS.840a2018S,2016MNRAS.462.2177K}. The calculation presented by Refs.~\cite{2016PhRvL.116w1102S,2017JPhCS.840a2018S} is essentially equivalent to our own calculation in Sec.~\ref{datapredictions}. Instead of explicitly considering a time window, those calculations randomize over the binary emission frequency (or, equivalently, $t_{\rm merger}$) and apply the SNR cut to each Monte Carlo sample.
In particular, Ref.~\cite{2017JPhCS.840a2018S} presents results using the same LISA noise curve used in this work, and quotes values of $N_{\rm obs}$ which are up to two orders of magnitude higher than ours. These differences can be reconciled 
by noting that Refs.~\cite{2016PhRvL.116w1102S,2017JPhCS.840a2018S} used the intrinsic merger rate $\mathcal{R}$ from Ref.~\cite{2016ApJ...833L...1A}, which is now superseded by Ref.~\cite{2018arXiv181112907T}.

\citet{2016MNRAS.462.2177K} presented a mostly analytical calculation of the expected number of stellar-mass BH binaries detections by LISA. They used the now-outdated LISA noise curves from Ref.~\cite{2016PhRvD..93b4003K}, but their N2A2 configuration with $T_{\rm obs}=5$~yr should be reasonably close to our calculations with $T_{\rm obs}=4$~yr (see also \cite{2017JPhCS.840a2018S}). They fix the chirp mass of all sources to $M_{c}=28 M_\odot$ and the intrinsic rate  to $\mathcal{R}=100$ Gpc${^{-3}}$yr${^{-1}}$ to obtain  $\ssim 30$  ($\rm 300$) observable sources that do (do not) merge within $T_{\rm obs}$. Their estimate is in rough agreement with our findings once $M_{c}$ and $\mathcal{R}$ are  properly rescaled.

Much of our analysis emphasized that LISA will be the limiting instrument of a multiband network. If a binary is detectable by LISA, it will also be observed by a ground-based observatory with high SNR (although it might take a very long time to merge). The sensitivity of the ground-based detector does not matter when estimating multiband detection rate. Differences between the current LIGO/Virgo network and future third-generation detectors impact the rate only for highly subthreshold events with $\rho_{\rm LISA} \lesssim 2$. Our predictions are presented assuming a ground-based network duty cycle $\mathcal{F}$ equal to $100\%$, thus envisioning future scenarios with multiple instruments operating concurrently. The merger time of multiband events can be predicted within $1\!-\!10$s \cite{2016PhRvL.116w1102S}, allowing to plan ground operations accordingly.

Of all stellar-mass BH binaries detected from the ground, LISA will select a specific subpopulation. In particular, sources must be more massive than $\ssim 60 M_\odot$ and closer than $\ssim 500$~Mpc, but the precise thresholds depend on the specifications of the LISA mission. 

In Fig.~\ref{massfilter} we compare our stellar calculations to posterior distributions from all LIGO/Virgo observations to date~\cite{2018arXiv181112907T}. Out of the 10 BH binary events observed so far, 5 have $z\lesssim 0.1$, 3 have $0.1\lesssim z \lesssim 0.2$, and 2 have $z\gtrsim 0.3$. Even with this limited sample, one can immediately see that expanding the LISA horizon redshift from $\ssim 0.1$ to $\ssim 0.2$ might dramatically increase our prospects of performing multiband observations.
Our more detailed analysis confirms this expectation. Because of the steep dependence of the detection rate on the LISA SNR, an extension of the LISA mission duration and/or data-analysis techniques that can lower the SNR threshold will increase the number of events by orders of magnitude.

The sharp high-mass cutoff in the multiband detectability might have other surprising consequences. In the populations considered here, the total mass of the binary is limited to $M\lesssim 100 M_\odot$ by construction. The distributions of Sec.~\ref{datapredictions} explicitly exclude BHs larger than $50 M_\odot$ \cite{2018arXiv181112907T}. 
The simulations used in Sec.~\ref{datapredictions} are generated assuming that pair-instability 
pulsation Supernovae and pair-instability Supernovae \cite{1964ApJS....9..201F,1967PhRvL..18..379B,2016A&A...594A..97B,2018arXiv181013412M,2017MNRAS.470.4739S} efficiently prevent the formation of BHs heavier than $\ssim 45 M_\odot$.
This  is a rather pessimistic assumption, as the cutoff might be as large as $\ssim 52 M_{\odot}$ \cite{2017ApJ...836..244W}. It is also
important to note that stars are predicted to form BHs with mass $\gtrsim 130 M_{\odot}$, for which Supernova instabilities are too weak to disrupt the progenitor star \cite{2017ApJ...836..244W}.
Second-generation mergers \cite{2017PhRvD..95l4046G,2018PhRvD..97l3003K} might also populate this upper mass gap. If such massive BHs exist, they are expected to contribute prominently to the multiband event rates~\cite{2010ApJ...722.1197A,2012PhRvD..85l3005K}.

If realized, multiband GW detections will crown the science return of the LISA mission with revolutionary astronomy, fundamental physics and cosmology. The analysis presented in this paper confirms that this is indeed an exciting possibility.

\acknowledgments

Data to reproduce results of this paper are publicly available at \href{https://github.com/dgerosa/spops}{github.com/dgerosa/spops} \cite{spops}. 
We thank Baoyi Chen, Ron Tso, Chris Moore, Antoine Klein, and Alberto Vecchio for discussions.
We thank~\citet{2018arXiv180301944R} for publicly sharing their codes to compute LISA SNRs and~\citet{2017arXiv170908079C} for publicly sharing their code to compute redshifted volumes, which was used for benchmarking.  Calculations of $p_{\rm det}$ are performed with the \textsc{gwdet}~\cite{gwdet} code which makes use of \textsc{pycbc}~\cite{2016CQGra..33u5004U} and \textsc{lal}~\cite{LAL}. The distributions used in Sec.~\ref{spopssection} are publicly available at~\cite{spops}, and are obtained with the \textsc{Startrack}~\cite{2008ApJS..174..223B} and \textsc{precession}~\cite{2016PhRvD..93l4066G} codes. 
D.G. is supported by NASA through Einstein Postdoctoral Fellowship Grant No. PF6-170152 awarded by the Chandra X-ray Center,  operated by the Smithsonian Astrophysical Observatory for NASA under Contract NAS8-03060.
E.B. and K.W.K.W. are supported by NSF Grant No. PHY-1841464, NSF Grant No. AST-1841358, NSF-XSEDE Grant No. PHY-090003, and NASA ATP Grant No. 17-ATP17-0225. 
R.O'S. is supported by NSF Grants No. PHY-1707965 and No. PHY-1607520.
This work has received funding from the European Union's H2020 research and innovation programme under the Marie Skłodowska-Curie Grant Agreement No. 690904. KB acknowledges partial support from the Polish National Science Center (NCN) 
grants OPUS (2015/19/B/ST9/01099) and Maestro (2015/18/A/ST9/00746). 
The authors would like to acknowledge networking support by the
European COST Action CA16104. %
Computational work was performed on Caltech cluster Wheeler supported by the Sherman Fairchild Foundation and Caltech, on the University of Birmingham's BlueBEAR cluster, and at the Maryland Advanced Research Computing Center (MARCC).
\appendix

\section{Additional results}
\label{suppmat}

In this Appendix we provide some of our rate calculations in tabular format. In particular, Table~\ref{tableallnumbers} lists results from the Monte Carlo runs of Sec.~\ref{datapredictions} for all choices of mass distribution, intrinsic merger rate and detector specifications. Table~\ref{startracktable} complements Sec.~\ref{spopssection} with detailed rates and number of observations from all of our population-synthesis simulations. It is worth stressing that the catalogs used to generate Table~\ref{startracktable} are produced by initializing zero-age-main-sequence stars at cosmological distances and by considering only BHs that merge before $z=0$. Many more {\em nonmerging} BH binaries are potentially detectable by LISA~\cite{2018MNRAS.480.2704L}, but they were not considered here.

\section{Horizon redshifts}
\label{zhorsec}

A prerequisite (and by-product) of the rate analysis presented in this paper is the calculation of the horizon redshift, defined as the largest redshift at which a binary with given parameters $\lambda$ is observable. More precisely, we define the horizon redshift $z_{\rm h}$ as the solution of the equations $\rho_{\rm opt}(\lambda, z_{\rm h})=\rho_{\rm thr}$ for ground-based detectors, and $\rho(t_{\rm merger}\!\!=\!\!T_{\rm obs},\lambda,z_h)=\rho_{\rm thr}$ for LISA.

\begin{figure*}[p]
\includegraphics[width=0.39\textwidth,page=5]{horizonredshift}
\includegraphics[width=0.39\textwidth,page=6]{horizonredshift}
\includegraphics[width=0.39\textwidth,page=1]{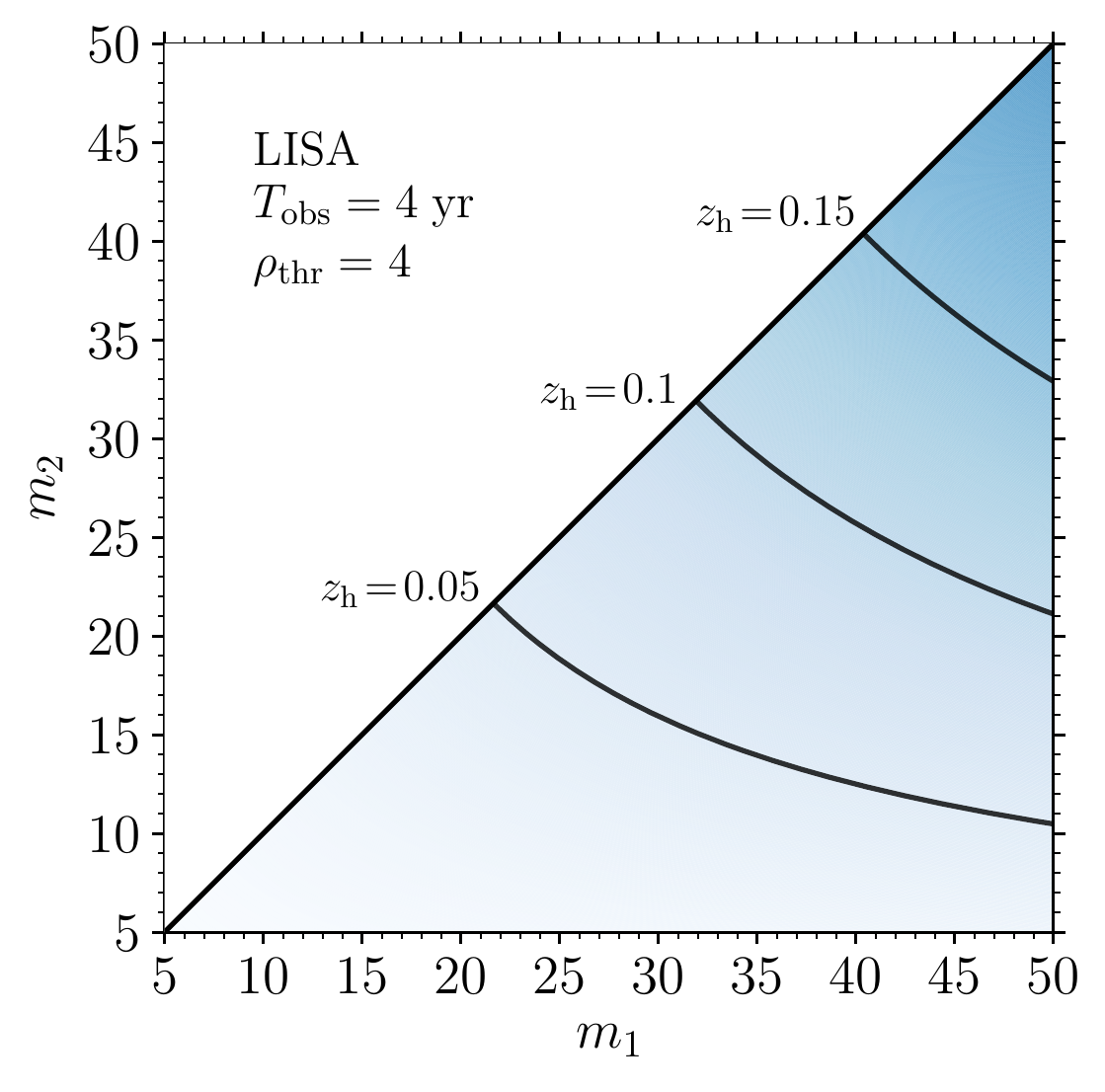}
\includegraphics[width=0.39\textwidth,page=2]{horizonredshift}
\includegraphics[width=0.39\textwidth,page=3]{horizonredshift}
\includegraphics[width=0.39\textwidth,page=4]{horizonredshift}

\caption{Stellar-mass BH binary horizon redshifts $z_{\rm h}$ for ground- and space-based GW detectors as a function of the BH masses $m_1>m_2$. Top panels are produced assuming LIGO at design sensitivity and the proposed third-generation detection Cosmic Explorer, together with the standard threshold $\rho_{\rm thr}=8$. Bottom and middle panels show results for the LISA space mission, assuming different mission duration of $T_{\rm obs}=4,\,10$~yr and SNR thresholds $\rho_{\rm thr} = 4,\,8$. For simplicity, here we assume nonspinning sources.
}
\label{horizonredshift}
\end{figure*}

Results for LIGO~\cite{2018LRR....21....3A} and  Cosmic Explorer~\cite{2017CQGra..34d4001A} are shown in the top panels of Fig.~\ref{horizonredshift}. Current detectors are most sensitive to binaries with $m_1+m_2\ssim 100 M_{\odot}$ and can reach $z_{\rm hor}\ssim 2$. The sensitivity of third-generation detectors, on the other hand, is expected to peak at lower masses $m_1+m_2\ssim 10 M_{\odot}$. Future interferometers will observe binaries out to $z_{\rm h}\gtrsim 30$, thus detecting all stellar-mass BH mergers in the Universe~\cite{2017CQGra..34d4001A,2010CQGra..27s4002P}. 

The middle and bottom panels of Fig.~\ref{horizonredshift} show forecasts for LISA, assuming different choices for the mission duration
 and SNR threshold.
For the stellar-mass BH binaries considered here, the LISA sensitivity peaks at the high end of the mass spectrum, where sources spend the longest time chirping in the detector's sensitivity window. With an extended mission duration of $T_{\rm obs}=10$~yr and an optimistic SNR threshold of  $\rho_{\rm thr}=4$, the furthest detectable binaries are located at $z_{\rm h}\ssim 0.25$. This is well below the reach of current and future ground-based detectors.

\begin{table*}[p]
\begin{tabularx}{\textwidth}{@{\hskip 10pt}l@{\hskip 12pt}l@{\hskip 10pt}l@{\hskip 15pt}l@{\hskip 15pt}||@{\hskip 10pt}c@{\hskip 10pt}c@{\hskip 10pt}c@{\hskip 7pt}|@{\hskip 7pt}c@{\hskip 10pt}c@{\hskip 10pt}c@{\hskip 5pt}X}
&&&& &\boldmath$r\; [{\rm yr}^{-1}]$& &  \multicolumn{3}{c}{\boldmath$N$} \\
&&&& Lower & Median & Upper & Lower & Median & Upper\\
\hline
\hline
\bf LISA & $T_{\rm obs}=10$ yr& $\rho_	{\rm LISA}>8$ & powerlaw & 1.18 & 2.11 & 3.59 & 11.84 & 21.09 & 35.88 \\
 &  &  & log & 1.70 & 2.99 & 5.03 & 16.99 & 29.89 & 50.33 \\
 & $T_{\rm obs}=10$ yr& $\rho_{\rm LISA}>4$ & powerlaw & (9.68) & (17.24) & (29.34) & (96.79) & (172.41) & (293.40) \\
 &  &  & log & (13.45) & (23.67) & (39.86) & (134.52) & (236.66) & (398.58) \\
 & $T_{\rm obs}=4$ yr& $\rho_{\rm LISA}>8$ & powerlaw & 0.76 & 1.36 & 2.31 & 3.04 & 5.42 & 9.23 \\
 &  &  & log & 1.08 & 1.90 & 3.19 & 4.31 & 7.58 & 12.77 \\
 & $T_{\rm obs}=4$ yr& $\rho_{\rm LISA}>4$ & powerlaw & (6.01) & (10.71) & (18.23) & (24.06) & (42.85) & (72.92) \\
 &  &  & log & (8.73) & (15.35) & (25.86) & (34.90) & (61.41) & (103.42) \\
\hline
\bf LISA+LIGO & $T_{\rm obs}=10$ yr& $\rho_{\rm LISA}>8$ & powerlaw & 0.47 & 0.83 & 1.42 & 4.69 & 8.35 & 14.21 \\
 &  &  & log & 0.74 & 1.31 & 2.20 & 7.44 & 13.09 & 22.05 \\
 & $T_{\rm obs}=10$ yr& $\rho_{\rm LISA}>4$ & powerlaw & 3.71 & 6.61 & 11.26 & 37.13 & 66.14 & 112.56 \\
 &  &  & log & 5.79 & 10.18 & 17.15 & 57.89 & 101.84 & 171.51 \\
 & $T_{\rm obs}=4$ yr& $\rho_{\rm LISA}>8$ & powerlaw & 0.18 & 0.33 & 0.56 & 0.73 & 1.30 & 2.22 \\
 &  &  & log & 0.29 & 0.51 & 0.87 & 1.17 & 2.06 & 3.46 \\
 & $T_{\rm obs}=4$ yr& $\rho_{\rm LISA}>4$ & powerlaw & 1.47 & 2.62 & 4.45 & 5.87 & 10.46 & 17.80 \\
 &  &  & log & 2.35 & 4.14 & 6.98 & 9.42 & 16.57 & 27.90 \\
\hline
\bf LISA+3g & $T_{\rm obs}=10$ yr& $\rho_{\rm LISA}>8$ & powerlaw & 0.48 & 0.85 & 1.44 & 4.75 & 8.47 & 14.41 \\
 &  &  & log & 0.76 & 1.33 & 2.24 & 7.55 & 13.29 & 22.38 \\
 & $T_{\rm obs}=10$ yr& $\rho_{\rm LISA}>4$ & powerlaw & 3.91 & 6.96 & 11.84 & 39.07 & 69.60 & 118.45 \\
 &  &  & log & 6.11 & 10.75 & 18.11 & 61.13 & 107.55 & 181.13 \\
 & $T_{\rm obs}=4$ yr& $\rho_{\rm LISA}>8$ & powerlaw & 0.18 & 0.33 & 0.56 & 0.74 & 1.32 & 2.24 \\
 &  &  & log & 0.29 & 0.52 & 0.87 & 1.18 & 2.07 & 3.49 \\
 & $T_{\rm obs}=4$ yr& $\rho_{\rm LISA}>4$ & powerlaw & 1.51 & 2.69 & 4.59 & 6.05 & 10.78 & 18.34 \\
 &  &  & log & 2.43 & 4.27 & 7.20 & 9.71 & 17.09 & 28.78 \\
\hline
\bf LIGO & $(z<0.3)$&  & powerlaw & 48.93 & 87.15 & 148.31 &&& \\
 &  &  & log & 31.51 & 55.43 & 93.35 &&& \\
\hline
\bf 3rd gen. & $(z<0.3)$&  & powerlaw & 205.23 & 365.56 & 622.09 &&& \\
 &  &  & log & 69.34 & 121.99 & 205.46 &&& \\
\end{tabularx}
\caption{Detection rates $r$ and number of observations $N$ inferred from current  LIGO/Virgo measurements of the intrinsic merger rate (cf. Sec.~\ref{datapredictions}). Results are reported for the LISA mission alone [Eqs. (\ref{pdetspaceonly})], as well as combined detections with the ground-based interferometers LIGO and Cosmic Explorer (``3g'') [Eq.~(\ref{pdetmultibib})]. For each estimate, we report three values (``lower'', ``median'', ``upper'') to bracket  uncertainties. Rates are estimated using the two populations of Ref.~~\cite{2018arXiv181112907T} (``powerlaw'' and ``log'') assuming both the nominal ($T_{\rm obs}$=4yr) and the extended ($T_{\rm obs}$=10yr) duration for the LISA mission. Events with $\rho_{\rm LISA}>8$ can be distinguished from space and will serve as predictions for the ground-based instruments. Events with $\rho_{\rm LISA}>4$ are not likely to be observed by LISA alone (hence the parentheses in the ``LISA'' entries of the table) but will be observable using ground-based detections as priors. For comparison, we also report the predicted event rate for LIGO at design sensitivity and Cosmic Explorer in the local Universe ($z<0.3$).}
\label{tableallnumbers}
\end{table*}

\begin{table*}[p]
\centering
\begin{tabularx}{\textwidth}{@{\hskip 5pt}l@{\hskip 15pt}l@{\hskip 10pt}l@{\hskip 8pt}||@{\hskip 8pt}c@{\hskip 10pt}c@{\hskip 10pt}c@{\hskip 10pt}c@{\hskip 10pt}c@{\hskip 10pt}c@{\hskip 10pt}cX}
 &&&&\multicolumn{5}{c}{{\bf Natal kick} \boldmath$\;\;\sigma$}  \\
$\;\;$\boldmath$r\; [{\rm yr}^{-1}]$ &&& $0$ km/s & $25$ km/s & $50$ km/s & $70$ km/s & $130$ km/s & $200$ km/s & $265$ km/s\\
\hline
\hline
\bf LISA & $T_{\rm obs}=10$ yr& $\rho_{\rm LISA}>8$ & 14.40 & 9.11 & 4.75 & 3.00 & 1.71 & 0.80 & 0.41 \\
 & $T_{\rm obs}=10$ yr& $\rho_{\rm LISA}>4$ & (124.25) & (78.34) & (40.12) & (25.36) & (13.39) & (6.16) & (3.62) \\
 & $T_{\rm obs}=4$ yr& $\rho_{\rm LISA}>8$ & 8.91 & 5.59 & 2.98 & 1.88 & 1.06 & 0.50 & 0.25 \\
 & $T_{\rm obs}=4$ yr& $\rho_{\rm LISA}>4$ & (74.52) & (47.24) & (24.45) & (15.39) & (8.64) & (4.01) & (2.16) \\
\hline
\bf LISA+LIGO & $T_{\rm obs}=10$ yr& $\rho_{\rm LISA}>8$ & 5.31 & 3.39 & 1.76 & 1.09 & 0.71 & 0.33 & 0.17 \\
 & $T_{\rm obs}=10$ yr& $\rho_{\rm LISA}>4$ & 48.68 & 30.20 & 15.49 & 9.65 & 5.21 & 2.38 & 1.59 \\
 & $T_{\rm obs}=4$ yr& $\rho_{\rm LISA}>8$ & 1.90 & 1.21 & 0.64 & 0.40 & 0.27 & 0.13 & 0.06 \\
 & $T_{\rm obs}=4$ yr& $\rho_{\rm LISA}>4$ & 17.08 & 10.72 & 5.55 & 3.38 & 2.10 & 0.97 & 0.60 \\
\hline
\bf LISA+3g & $T_{\rm obs}=10$ yr& $\rho_{\rm LISA}>8$ & 5.36 & 3.43 & 1.78 & 1.10 & 0.71 & 0.34 & 0.18 \\
 & $T_{\rm obs}=10$ yr& $\rho_{\rm LISA}>4$ & 50.63 & 31.40 & 16.15 & 10.06 & 5.43 & 2.49 & 1.67 \\
 & $T_{\rm obs}=4$ yr& $\rho_{\rm LISA}>8$ & 1.91 & 1.22 & 0.64 & 0.40 & 0.28 & 0.13 & 0.06 \\
 & $T_{\rm obs}=4$ yr& $\rho_{\rm LISA}>4$ & 17.46 & 10.96 & 5.69 & 3.46 & 2.15 & 1.00 & 0.61 \\
\hline
\bf LIGO &  &  & $ 3.2\!\times\! 10^3$ & $ 1.9\!\times\! 10^3$ & $ 8.9\!\times\! 10^2$ & $ 5.6\!\times\! 10^2$ & $ 2.4\!\times\! 10^2$ & $ 1.1\!\times\! 10^2$ & $ 6.4\!\times\! 10^1$ \\
\hline
\bf 3rd gen. &  &  & $ 9.4\!\times\! 10^5$ & $ 7.7\!\times\! 10^5$ & $ 4.4\!\times\! 10^5$ & $ 2.9\!\times\! 10^5$ & $ 1.3\!\times\! 10^5$ & $ 6.4\!\times\! 10^4$ & $ 3.8\!\times\! 10^4$ \\
&\\
&\\
 &&&&\multicolumn{5}{c}{{\bf Natal kick} \boldmath$\;\;\sigma$}  \\
\multicolumn{2}{l}{$\;\;$\boldmath$N$}&& $0$ km/s & $25$ km/s & $50$ km/s & $70$ km/s & $130$ km/s & $200$ km/s & $265$ km/s\\
\hline
\hline
\bf LISA & $T_{\rm obs}=10$ yr& $\rho_{\rm LISA}>8$ & 143.96 & 91.12 & 47.54 & 30.03 & 17.06 & 7.97 & 4.08 \\
 & $T_{\rm obs}=10$ yr& $\rho_{\rm LISA}>4$ & (1242.46) & (783.39) & (401.24) & (253.59) & (133.94) & (61.58) & (36.23) \\
 & $T_{\rm obs}=4$ yr& $\rho_{\rm LISA}>8$ & 35.65 & 22.35 & 11.93 & 7.52 & 4.25 & 2.01 & 0.99 \\
 & $T_{\rm obs}=4$ yr& $\rho_{\rm LISA}>4$ & (298.09) & (188.94) & (97.79) & (61.57) & (34.55) & (16.06) & (8.64) \\
\hline
\bf LISA+LIGO & $T_{\rm obs}=10$ yr& $\rho_{\rm LISA}>8$ & 53.10 & 33.95 & 17.61 & 10.90 & 7.05 & 3.32 & 1.75 \\
 & $T_{\rm obs}=10$ yr& $\rho_{\rm LISA}>4$ & 486.80 & 302.03 & 154.89 & 96.51 & 52.07 & 23.81 & 15.94 \\
 & $T_{\rm obs}=4$ yr& $\rho_{\rm LISA}>8$ & 7.58 & 4.85 & 2.55 & 1.59 & 1.10 & 0.52 & 0.26 \\
 & $T_{\rm obs}=4$ yr& $\rho_{\rm LISA}>4$ & 68.32 & 42.89 & 22.21 & 13.51 & 8.39 & 3.89 & 2.38 \\
\hline
\bf LISA+3g & $T_{\rm obs}=10$ yr& $\rho_{\rm LISA}>8$ & 53.64 & 34.30 & 17.80 & 11.02 & 7.14 & 3.36 & 1.77 \\
 & $T_{\rm obs}=10$ yr& $\rho_{\rm LISA}>4$ & 506.34 & 314.05 & 161.47 & 100.61 & 54.31 & 24.85 & 16.74 \\
 & $T_{\rm obs}=4$ yr& $\rho_{\rm LISA}>8$ & 7.62 & 4.88 & 2.57 & 1.59 & 1.11 & 0.52 & 0.26 \\
 & $T_{\rm obs}=4$ yr& $\rho_{\rm LISA}>4$ & 69.84 & 43.85 & 22.74 & 13.82 & 8.60 & 3.98 & 2.44 \\
\end{tabularx}
\caption{Detection rates ($r$, top) and number of events ($N$, bottom) for ground-based detectors (LIGO, Cosmic Explorer ``3g''), space missions (LISA) and multiband scenarios as predicted by population synthesis simulations of binary stars (cf. Sec.~\ref{spopssection}). We present results from 7 simulations, where we only vary the magnitude of kicks imparted to BHs at birth ($\sigma= 0$, 25, 50, 70, 130, 200, 265 km/s). We consider two different LISA mission durations $T_{\rm obs}=4,10$ yr and SNR thresholds $\rho_{\rm LISA}>8,4$. Events with $\rho_{\rm LISA}>4$ can only be extracted using ground-based data as priors, and are therefore indicated in parenthesis for LISA alone.}
\label{startracktable}
\end{table*}

\clearpage

\bibliography{multibandrates}

\end{document}